\documentclass[11pt]{emulateapj}
\usepackage{graphicx,amssymb,amsmath,times}
\usepackage{natbib,graphicx}
\usepackage[a4paper]{hyperref}
\newcommand {\apgt} {\ {\raise-.5ex\hbox{$\buildrel>\over\sim$}}\ }
\newcommand {\aplt} {\ {\raise-.5ex\hbox{$\buildrel<\over\sim$}}\ }

\newcommand{\flux}{\ensuremath{\mathrm{~erg~cm^{-2}~s^{-1}} }}

\def\flux{erg cm$^{-2}$ s$^{-1}$}
\def\latflux{$\times 10^{-6}$ ph cm$^{-2}$ s$^{-1}$}
\begin{document}
%
\title{PKS 1502+106: a new and distant gamma-ray blazar in outburst\\
discovered by the Fermi Large Area Telescope}
%
%
\shorttitle{PKS 1502+106: a new and distant gamma-ray blazar in
outburst discovered by the Fermi LAT}
%
%
\shortauthors{Abdo et~al.}

\author{
A.~A.~Abdo\altaffilmark{2,3},
M.~Ackermann\altaffilmark{4},
M.~Ajello\altaffilmark{4},
W.~B.~Atwood\altaffilmark{5},
M.~Axelsson\altaffilmark{6,7},
L.~Baldini\altaffilmark{8},
J.~Ballet\altaffilmark{9},
G.~Barbiellini\altaffilmark{10,11},
D.~Bastieri\altaffilmark{12,13},
B.~M.~Baughman\altaffilmark{14}
% ...
$~$\\
(The Fermi LAT Collaboration),\\
and \\
E.~Angelakis\altaffilmark{34},
T.~Hovatta\altaffilmark{62},
E.~Hoversten\altaffilmark{36},
Y.~Ikejiri\altaffilmark{35},
K.~S.~Kawabata\altaffilmark{63},
Y.~Y.~Kovalev\altaffilmark{64,34},
Yu.~A.~Kovalev\altaffilmark{64},
T.~P.~Krichbaum\altaffilmark{34},
M.~L.~Lister\altaffilmark{65},
A.~L\"ahteenm\"aki\altaffilmark{62},
N.~Marchili\altaffilmark{34},
P.~Ogle\altaffilmark{46},
C.~Pagani\altaffilmark{36},
A.~B.~Pushkarev\altaffilmark{66,34,67},
K.~Sakimoto\altaffilmark{35},
M.~Sasada\altaffilmark{35},
M.~Tornikoski\altaffilmark{62},
M.~Uemura\altaffilmark{63},
M.~Yamanaka\altaffilmark{35},
T.~Yamashita\altaffilmark{63}
} 
%
%
%
%
%
%
%
\small
% ...
\altaffiltext{2}{Space Science Division, Naval Research Laboratory, Washington, DC 20375, USA}
\altaffiltext{3}{National Research Council Research Associate, National Academy of Sciences, Washington, DC 20001, USA}
\altaffiltext{4}{W. W. Hansen Experimental Physics Laboratory, Kavli Institute for Particle Astrophysics and Cosmology, Department of Physics and SLAC National Accelerator Laboratory, Stanford University, Stanford, CA 94305, USA}
\altaffiltext{5}{Santa Cruz Institute for Particle Physics, Department of Physics and Department of Astronomy and Astrophysics, University of California at Santa Cruz, Santa Cruz, CA 95064, USA}
\altaffiltext{6}{Department of Astronomy, Stockholm University, SE-106 91 Stockholm, Sweden}
\altaffiltext{7}{The Oskar Klein Centre for Cosmoparticle Physics, AlbaNova, SE-106 91 Stockholm, Sweden}
\altaffiltext{8}{Istituto Nazionale di Fisica Nucleare, Sezione di Pisa, I-56127 Pisa, Italy}
\altaffiltext{9}{Laboratoire AIM, CEA-IRFU/CNRS/Universit\'e Paris Diderot, Service d'Astrophysique, CEA Saclay, 91191 Gif sur Yvette, France}
\altaffiltext{10}{Istituto Nazionale di Fisica Nucleare, Sezione di Trieste, I-34127 Trieste, Italy}
\altaffiltext{11}{Dipartimento di Fisica, Universit\`a di Trieste, I-34127 Trieste, Italy}
\altaffiltext{12}{Istituto Nazionale di Fisica Nucleare, Sezione di Padova, I-35131 Padova, Italy}
\altaffiltext{13}{Dipartimento di Fisica ``G. Galilei", Universit\`a di Padova, I-35131 Padova, Italy}
\altaffiltext{14}{Department of Physics, Center for Cosmology and Astro-Particle Physics, The Ohio State University, Columbus, OH 43210, USA}
% ...
\altaffiltext{34}{Max-Planck-Institut f\"ur Radioastronomie, Auf dem H\"ugel 69, 53121 Bonn, Germany}
\altaffiltext{35}{Department of Physical Sciences, Hiroshima University, Higashi-Hiroshima, Hiroshima 739-8526, Japan}
\altaffiltext{36}{Department of Astronomy and Astrophysics, Pennsylvania State University, University Park, PA 16802, USA}
% ...
\altaffiltext{46}{Cahill Center for Astronomy and Astrophysics, California Institute of Technology, Pasadena, CA 91125, USA}
% ...
\altaffiltext{62}{Mets\"ahovi Radio Observatory, Helsinki University of Technology TKK, FIN-02540 Kylmala, Finland}
\altaffiltext{63}{Hiroshima Astrophysical Science Center, Hiroshima University, Higashi-Hiroshima, Hiroshima 739-8526, Japan}
\altaffiltext{64}{Astro Space Center of the Lebedev Physical Institute, 117810 Moscow, Russia}
\altaffiltext{65}{Department of Physics, Purdue University, West Lafayette, IN 47907, USA}
\altaffiltext{66}{Crimean Astrophysical Observatory, 98409 Nauchny, Crimea, Ukraine}
\altaffiltext{67}{Pulkovo Observatory, 196140 St. Petersburg, Russia}
% ...
\normalsize
%
\begin{abstract}
%
%
The Large Area Telescope (LAT) on board the \textit{Fermi} Gamma-ray
Space Telescope discovered a rapid ($\sim$ 5 days duration), high-energy ($E >100$ MeV) gamma-ray outburst from a source identified with the blazar PKS 1502+106 (OR 103, S3 1502+10, z=1.839) starting on August 05, 2008 ($\sim 23$ UTC, MJD 54683.95), and followed by bright and variable flux over the next few months. Results on the gamma-ray localization and identification, as well as spectral and temporal behavior during the first months of the \textit{Fermi} all-sky survey are reported here in conjunction with a multi-waveband characterization as a result of one of the first \textit{Fermi} multi-frequency
campaigns. The campaign included a \textit{Swift} ToO (followed up by 16-day observations on August 07-22, MJD 54685-54700), VLBA (within the MOJAVE program), Owens Valley (OVRO) 40m, Effelsberg-100m, Mets\"{a}hovi-14m, RATAN-600 and Kanata-Hiroshima radio/optical observations. Results from the analysis of archival
observations by INTEGRAL, XMM-\textit{Newton} and \textit{Spitzer} space telescopes are reported for a more complete picture of this new gamma-ray blazar.
PKS 1502+106 is a sub-GeV peaked, powerful flat spectrum radio quasar (luminosity at $E>100$ MeV, $L_{\gamma}$, is about $1.1 \times 10^{49}$ erg s$^{-1}$, and black hole mass likely close to $10^{9}$ M$_\sun$), exhibiting marked gamma-ray bolometric dominance, in particular during the asymmetric outburst ($L_{\gamma}/L_{opt} \sim 100$, and 5-day averaged flux F$_{E>100~\mathrm{MeV}}=2.91\pm 1.4 \times 10^{-6}$ ph cm$^{-2}$ s$^{-1}$), which was characterized by a factor greater than 3 of flux increase in less than 12 hours.
The outburst was observed simultaneously from optical to X-ray bands (F$_{0.3-10~\mathrm{keV}}=2.18_{-0.12}^{+0.15} \times 10^{-12}$ \flux, and hard photon index $\sim 1.5$, similar to past values) with a flux increase of less than one order of magnitude with respect to past observations, and was likely controlled by Comptonization of external-jet photons produced in the broad line region (BLR) in the gamma-ray band. No evidence of a possible blue bump signature was observed in the optical-UV continuum spectrum, while some hints for a possible 4-day time-lag with respect to the gamma-ray flare were found. Nonetheless, the properties of PKS 1502+106 and the strict optical/UV, X- and gamma-ray cross correlations suggest the contribution of the synchrotron self Compton (SSC), in-jet, process should dominate from radio to X-rays. This mechanism may also be responsible for the consistent gamma-ray variability observed by the LAT on longer timescales, after the ignition of activity at these energies provided by the BLR-dissipated outburst. Modulations and subsequent minor, rapid flare events were detected, with a general fluctuation mode between pink-noise and a random-walk. The averaged gamma-ray spectrum showed a deviation from a simple power-law, and can be described by a log-parabola curved model peaking around 0.4-0.5 GeV. The maximum energy of photons detected from the source in the first four months of LAT observations was $15.8$ GeV, with no significant consequences on extragalactic background light predictions. A possible radio counterpart of the gamma-ray outburst can be assumed only if a delay of more than 3 months is considered on the basis of opacity effects at cm and longer wavelengths. The rotation of the electric vector position angle observed by VLBA from 2007 to 2008 could represent a slow field ordering and alignment with respect to the jet axis, likely a precursor feature of the ejection of a superluminal radio knot and the high-energy outburst. This observing campaign provides more insight into the connection between MeV-GeV flares and the moving, polarized structures observed by the VLBI.
\end{abstract}


%
%
\section{Introduction}\label{sect:introduction}
%
%
The Large Area Telescope (LAT), on board the \textit{Fermi} Gamma-ray
Space Telescope \citep[formerly GLAST; ][]{ritz07}, was successfully
launched by NASA on 2008, June 11, from Cape Canaveral, Florida, on a Delta II Heavy launch vehicle.  While still in the commissioning and checkout phase,
it discovered and monitored bright, flaring gamma-ray emission above
100 MeV from a source identified with the blazar PKS 1502+106
(historically also known as OR 103 and S3 1502+10).  The large field of view, effective area and sensitivity and the nominal survey observational mode make \textit{Fermi}-LAT an unprecedented all-sky monitor of $\gamma$-ray flares and source variability \citep[see, e.g. ][]{mcenery06,thompson06,lott07,michelsonLAT08}.
\par At the beginning of August 2008, PKS 1502+106 was the second
brightest extragalactic source in the $\gamma$-ray
sky, exhibiting a sudden high-energy outburst announced in ATel \#1650.  This outburst
successfully triggered the first (unplanned) \textit{Fermi} multi-frequency campaign.
Major renewed gamma-ray activity observed by \textit{Fermi} in January 2009 was announced via ATel \#1905.

\par PKS 1502+106 is a luminous, quasar-like (optically broad-line and flat radio
spectrum) AGN discovered during the 178 MHz pencil beam survey from
the Mullard Radio Astronomy Observatory, Cambridge, UK, \citep[appearing in a
list not included in the 4C catalog; ][]{crowter66,williams67},
and was re-observed and characterized as an extragalactic
source by both the Australian National Radio Astronomy Observatory
of Parkes, NSW, Australia, \citep[][ id.: PKS 1502+106]{day66},
and the Ohio State University (``Big Ear'') Radio Observatory,
Delaware, OH, USA, \citep[][ id.: OR 103]{fitch69}. The source
exhibited substantial radio flux variations (factor $>2$), a high
degree of linear polarization, a core-dominated, one-sided and
curved radio jet with a variable, a complex morphology at VLBI
scales \citep{an04,lister09a}, and a compact large scale
structure. 11 VLBA observations at 15.4 GHz performed between Aug.~1997 and Aug.~2007 showed a FWHM major beam axis in the range 1.02-1.57 mas, a minor axis beam axis of 0.5 mas, and a total flux density in the range 0.88-1.93 Jy \citep{lister09a}, and apparent jet speed  of $(14.8 \pm 1.2)c$ \citep{lister09b}. The 22 and 37 GHz flux history shows several long-term flares ($>$ 1 Jy variations, i.e. $\sim 60\%$ of the total flux range span, on typical timescales of a year, and peak fluxes well above 2 Jy), with at least five flares and an average trend that was slightly increasing from 1988 to mid-2004 \citep{terasranta05}. WMAP fluxes at similar frequencies (K, Ka, Q bands) are in agreement with these flux ranges \citep{lopez07}. The Doppler factor estimated from the observed 37 GHz variability and brightness temperature \citep{hovatta09} agrees with the jet speed ($14.6c$) cited above, a Doppler factor $\mathcal{D}_{var}=12$ and viewing angle $\theta_{var}=4.7^{\circ}$.

\par This radio blazar was identified in the
optical band by \citet{blake70} with a position refinement by
\citet{argue80}, while an initial spectroscopic inspection was performed
by \citet{burbidge72}. Variations $>2.5$ mag were observed in its
optical flux history (Palomar-Quest and Catalina Sky Surveys, ATel
\#1661), together with a variable and relatively high degree (up to
$20\%$) of linear polarization, pointing out a dominant
synchrotron emission with no observed dilution by thermal
components. The redshift of PKS 1502+106, as
confirmed by the good S/N spectrum of the Sloan Digital Sky Survey
($z=1.8385\pm 0.0024$ at high confidence), is in agreement with
the value $z=1.839$ estimated previously by \citet{smith77}.
A less remote value ($z=0.56$) is
reported in other works \citep{burbidge72,wright79,wilkes83}, although
the possible multiple MgII absorption system (pointed
out by a feature shortward of the 4388 \AA$~$ emission
line) would be very unusual for a low redshift object.
\par Serendipitous X-ray data of PKS 1502+106 are available because
the source lies about $7'$ NE of the bright Seyfert type-1 galaxy Mkn
841, although only one multifrequency work dedicated
to this blazar \citep{george94} has appeared previously. Early X-ray observations (ROSAT, ASCA) showed
low-amplitude variations on short timescales (factor $> 2$ on timescales of a year),
a flat 0.1-10 keV photon index $\Gamma_{X}$ between
1.4 and 1.9, and an intrinsic X-ray luminosity of
$L_{2-10\mathrm{keV}}=1.2 \times 10^{46}$ erg s$^{-1}$, and a 2-10 keV flux in the range $4.9-6.54 \times 10^{-13}$ \flux \citep{george94,akiyama03,watanabe04}. PKS 1502+106 was speculated
to be a possible $\gamma$-ray source before the LAT discovery because of the superluminal motions of jet components \citep[up to $187\pm 15
\mu$as/year][]{lister05,an04}, and the multiwaveband spectral
indexes $\alpha_{rx}$ and $\alpha_{ox}$ \citep[consistent with
other FSRQs detected by EGRET, ][]{george94}. Only modest intrinsic X-ray absorption was suggested by this work, and the optical and near-IR reddening
claimed in \citet[][]{watanabe04} is probably due to the synchrotron jet dominance at these low frequencies rather than by absorption from inner nuclear light.
\par A relation involving the misalignment between the pc- and kpc-scale radio structure (position angle) and the $\gamma$-ray emission was postulated as
well \citep{cooper07}. However, only a cumulative $2\sigma$ upper
limit by EGRET  of $7\times 10^{-8}$ ph cm$^{-2}$ s$^{-1}$ was
reported \citep[Phase/Cycle I, combined Viewing Periods: 24.0 to
25.0, i.e.~April 02-23, 1992; ][]{fichtel94}, and the source was likewise
undetected in the following EGRET cycles \citep{hartman99,casandjian08}.
\par In the following we use a $\Lambda$CDM (concordance) cosmology with values given within 1$\sigma$ of the
WMAP results \citep{komatsu08}, namely $h=0.71$, $\Omega_{m}=0.27$
and $\Omega_{\Lambda}=0.73$, and a Hubble constant value $H_{0}=100h$
km s$^{-1}$ Mpc$^{-1}$.
\par In Section 2, first results on the $\gamma$-ray identification,
the observed MeV-GeV outburst and the subsequent four months of monitoring by the \textit{Fermi}-LAT are described. In Section 3, multifrequency results obtained through simultaneous optical-UV-X-ray observations by \textit{Swift} (thanks to a 16-day long monitoring following a triggered Target of Opportunity, ToO), and by radio-optical observatories (the 40m dish
telescope of the Owens Valley Radio Observatory, the Effelsberg 100m
dish radio telescope, the ring radio telescope RATAN-600, the VLBA within the MOJAVE program, and the Kanata telescope of the Higashi-Hiroshima Observatory) are summarized. In addition, past and unpublished observations by the XMM-Newton and Spitzer space telescopes are analyzed and presented in Section 4 for a more complete picture. Finally, in Section 5 and 6, discussion and concluding remarks are reported.
%
%
\section{Gamma-ray observations and results by \textit{Fermi}-LAT}\label{sect:LAT}
%
%
\subsection{LAT observations}\label{subsect:latobservation}
%
%
The LAT instrument is a pair
tracker-converter telescope comprising a modular array of 16
towers---each with a tracker based on silicon micro-strip detector
technology---and a calorimeter based on a hodoscopic array of 96 CsI(Tl) crystals, surrounded by an Anti-Coincidence
Detector capable of measuring the directions and energies of cosmic $\gamma$-ray photons with energies from 20 MeV to $>300$ GeV \citep[for details, see, e.g. ][]{bellazzini02,michelson07,atwood07,michelsonLAT08,lat-calibration}.
%
%
\begin{figure}[tt!!]
\centering 
\resizebox{\hsize}{!}{\rotatebox[]{0}{\includegraphics{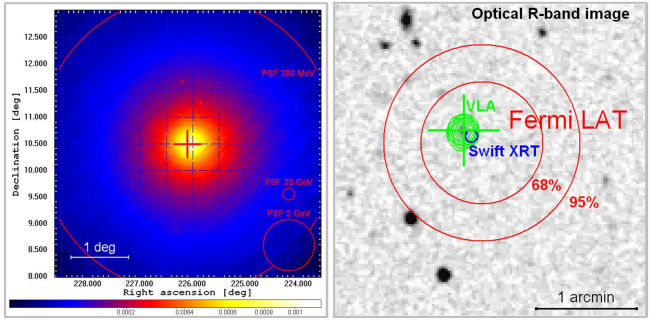}}}
\vskip 0.1cm \caption{
\textit{Left panel}: LAT count map cumulated on a nine-month (Aug. 2008 - Apr. 2009) baseline, weighted and smoothed
by the point spread function (PSF) such that higher energy photons are
mapped to higher intensities. The map is in arbitrary units in the energy
range 0.1-100 GeV and in a $2.5^{\circ} \times 2.5^{\circ} $ region centered
on PKS 1502+106. The qualitative circle sizes of the PSF at 200MeV, 2GeV and 20GeV are outlined for reference.
\textit{Right panel}: LAT source localization with 95\% and 68\% uncertainty radii (red circles)
superimposed on an arcmin-scale (R-band) optical image showing also the
X-ray counterpart error box by the \textit{Swift}-XRT observations and the radio
position and intensity contours by VLA of PKS 1502+106. The best LAT source
position, calculated on the same nine-month period, with the \texttt{pointlike} tool is RA: 226.10179$^{\circ}$, Dec:
+10.4927$^{\circ}$, $\Delta = 0.0027^{\circ}$, with 68\% and 95\% LAT error
circles of $0.0077^{\circ}$ and $0.0124^{\circ}$ respectively.
} \label{fig:countmap}
\end{figure}
%
%
\par The reduction and analysis of LAT data was performed using the
Science Tools v.9.8, based in particular on a unbinned maximum-likelihood estimator
of the spectral model parameters (\texttt{gtlike} tool). Events were
selected using the Instrument Response Functions (IRFs) P6\_V1\_DIFFUSE.  This selection provides the cleanest set of events (in terms of
direction, energy reconstruction and background rejection) at
the cost of reduced effective area at low energies, and takes
into account the differences between front- and back-converting events.
To minimize contamination by Earth albedo $\gamma$-ray events
that have reconstructed directions with angles with respect to the
local zenith $>105^{\circ}$ have been excluded.
For this object with high Galactic latitude, events are
extracted within a $10^\circ$ acceptance cone centered at the PKS
1502+106 radio position. This cone, substantially larger than the 68\% containment
angle of the PSF at the lowest energies, provides sufficient events
to accurately constrain the diffuse emission components. The \texttt{gtlike} model includes the PKS 1502+106 point source component, two other point sources from the 3 month catalog (both faint and low-confidence sources with TS $\simeq 0.9\%$ of the TS value of PKS 1502+106 for the same period), a component for the Galactic diffuse emission \citep[GALPROP code, see, e.g.
][ and references therein]{moskalenko03}, and an isotropic
component including the extragalactic diffuse
emission and the residual background from cosmic rays.

\par The \textit{Fermi}-LAT data of PKS 1502+106 presented here were obtained during the first four months of the LAT survey (Aug.-Dec. 2008). In this period PKS 1502+106 was one of the most persistently bright, variable sources in the high-energy sky and almost certainly the source with the highest luminosity. The background contribution within a few degrees was only a small fraction of the source count rate, with no nearby source confusion. The time interval was sufficient for a fine determination of the average spectrum, for a first look at the mid-timescale variability and detection of posterior flares, for a refined localization, and a first cross comparison with the other multifrequency monitoring data. The first $\gamma$-ray detection of PKS 1502+106 by the LAT occurred in July 2008, when it was confirmed by the high-level Automatic Science Processing pipeline monitoring
\citep[ASP; ][]{chiang06,chiang07}, based on a wavelet-based (\texttt{pgwave}) quick-look detection tool \citep[e.g. ][]{damiani97,marcucci04,ciprini07a} and a maximum likelihood analysis, and by the LAT Source Catalog algorithm (\texttt{mr\_filter}) based on wavelet analysis in the Poisson regime \citep{starck98} and the peak-finding tool \texttt{sExtractor} \citep{bertin96}.
The rapid and markedly time-asymmetric $\gamma$-ray outburst announced in ATel \#1650 and triggering an unplanned Target of Opportunity (ToO) multifrequency campaign was seen from Aug.~05 (about 23 UTC) until about Aug.~11, 2008 ($\sim 5$ days duration), with a fast rise, slower decay, and an approximately two-day sustained peak flux.
\par Some caveats related to the preflight instrument response functions (P6\_V1), which overestimated the acceptance at low energies, are briefly described in \citet{3monthsagn}.
%
%
%
\begin{figure*}[tbh!!!]
\centering %
\hspace{-0.5cm} %
\resizebox{\hsize}{!}{\rotatebox[]{90}{\includegraphics{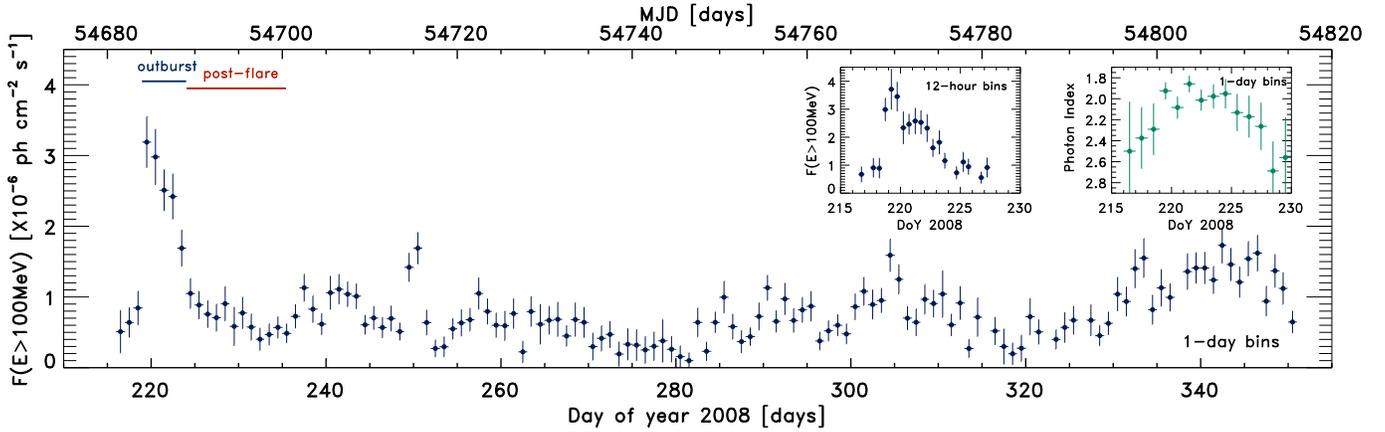}}}%
\vspace{-6.2cm} \caption{\textit{Main panel:} Likelihood flux (E$>$100MeV) light curve
obtained in daily bins from Aug. 02 to Dec. 15, 2008. The outburst state and the
subsequent post-flare (a lower and intermediate level brightness, far from the faintest state observed) period with
simultaneous monitoring by \textit{Swift} are represented by the two horizontal
lines. \textit{Left inset plot:} A zoom on the corresponding flux light
curve around the outburst period obtained using finer, 12-hour bins (lower statistics). \textit{Right inset plot:} The gamma-ray (E$>$100MeV) photon index values for the same period using daily bins as in the main panel light curve.} \label{fig:LATlightcurve}
\end{figure*}
%
%
\subsection{Gamma-ray source localization, association and identification}\label{subsec:association}
%
%
The LAT PSF and sensitivity provides an unprecedented
angular resolution in gamma-rays \citep[68\% containment radius
better than $\sim 1^\circ$ at 1 GeV,][]{atwood07,cecchi07,michelsonLAT08,3monthcatalog,lat-calibration},
making the association and identification processes less difficult
than in previous experiments. In the case of this very bright $\gamma$-ray source, we obtained---beyond the good spatial ``association''---a firm ``identification''
with PKS 1502+106. The 3-month bright source list results \citep[][id: 0FGL J1504.4+1030]{3monthcatalog}, provided a good initial localization: RA: 226.12$^{\circ}$, Dec: +10.51$^{\circ}$; $r_{95}=0.05^{\circ}$ and $\sqrt{TS}=88.2$, ($r_{95}$ being the radius of 95\% confidence region, TS the likelihood test statistic from the 200 MeV to 100 GeV analysis). Application of the \texttt{pointlike} tool \citep{burnett07,3monthcatalog} on a much longer (nine-month, Aug.~2008 - Apr.~2009)
LAT dataset with very high statistics, provided an excellent localization (outlined in Fig.\ref{fig:countmap}):
RA: 226.10179$^{\circ}$, Dec: +10.4927$^{\circ}$, $\Delta = 0.0027^{\circ}$, with 68\% and 95\% LAT error
circles of $0.0077^{\circ}$ and $0.0124^{\circ}$ respectively (statistical only). Studies of bright source localizations indicate a systematic uncertainty in the localization of $<30$'', that can be taken as an estimate of the
systematics with this tool \citep[more details on the production of \texttt{pointlike} density maps and localization are described in ][]{camilo09}. The relevant improvement in the localization carried out on a nine-month baseline is due to the high variability that occurred with this source. These localization values are in agreement with the VLBI radio and optical positions of PKS 1502+106, the VLA contours and the \textit{Swift} XRT error box (Fig. \ref{fig:countmap}). PKS 1502+106 is the only bright VLA radio source (calibrator source list) located within the LAT 95\% confidence circle. The Seyfert galaxy Mkn 841 (observed with a hard X-ray cutoff, see Sect.\ref{sect:otherMW}) is positioned well
outside of these localization circles, as are other HB89-catalog quasars in this region.
\par Beyond the excellent spatial association, the most secure and distinctive signature for firm identification of this new gamma-ray source found by \textit{Fermi} is the observed correlation between the $\gamma$-ray, X-ray and optical-UV variability (see sections \ref{sect:simultaneousmwobs} and \ref{sect:otherMW}). This object was also a member of the pre-launch CGRaBS \citep[][object id: CGRaBS J1504+1029]{healey08} and Roma-BZCAT \citep[][object id: BZQ J1504+1029 ]{massaro08} catalogs listing candidate gamma-ray blazars. Finally, a method based on a ``figure of merit'' \citep[described in ][]{sowards03,sowards05} for this LAT source position provides a very high likelihood of identification with PKS 1502+106.
%
%
%
%
%
\subsection{Gamma-ray temporal behavior}\label{subsect:gamma-lightcurve}
%
%

The typically bright $\gamma $-ray flux and the enduring activity shown by PKS 1502+106 in $\gamma$-rays, allowed a firm detection of the source by the LAT on a daily basis.
Fig. \ref{fig:LATlightcurve} shows the light curve (daily bins, $E>100$MeV) extracted with the \texttt{gtlike} tool over the first four and half months of LAT all-sky survey. A fast-rising, markedly asymmetric and bright outburst was found, with a factor $> 3$ of increase in flux in less than 12 hours. The integrated flux at $E>100$ MeV averaged in the 12h bin of the peak emission (between Aug. 05 and Aug. 06, 2008, i.e.  DOY 218-219) was $F_{E>100~\mathrm{MeV}} = (3.7\pm0.7)\times 10^{-6}$ ph cm$^{-2}$ s$^{-1}$ (statistical only), as measured when the LAT instrument was still in commissioning and checkout phase (all-sky nominal mode). The emission from PKS 1502+106 then faded more slowly in the
following days, and the entire outburst interval spanned Aug.~05 around 23 UTC until to Aug.~11 around 00 UTC, 2008 (i.e. DOY, 218.95-224.0, $\sim 5$ days duration Fig. \ref{fig:LATlightcurve}). The peak flux appeared elevated for less than two days, rivaling the brighest apparent flux of other extragalactic sources at that time (Section \ref{subsect:latobservation}). The finer, 12-hour bin light curve ($\sim 8$ \textit{Fermi} orbits, ensuring still similar exposures) reported in the right inset panel of Fig. \ref{fig:LATlightcurve}, shows the elevated flux held for at least one 1.5 days, while the slower fading decay exhibits a high flux ($F_{E>100~\mathrm{MeV}} > 2$ \latflux) plateau, or a secondary superposed flare, that extended for about 2.5 days. During this outburst a ``harder when brighter'' spectral trend is suggested, despite the photon index error dispersion (see the daily photon indices reported on the right inset panel of Fig. \ref{fig:LATlightcurve}). The \texttt{gtlike} performances and the current IRFs used did not allow to go below a daily binning in the extrapolation of the photon index values, and this was possible in an acceptable way with respect to amplitude of the statistical error only for the high-flux and high photon count statistics available around the outburst epoch.
%
%

\begin{figure}[tt!!!]
\centering %
\hspace{-0.1cm} %
\resizebox{7.5cm}{!}{\rotatebox[]{0}{\includegraphics{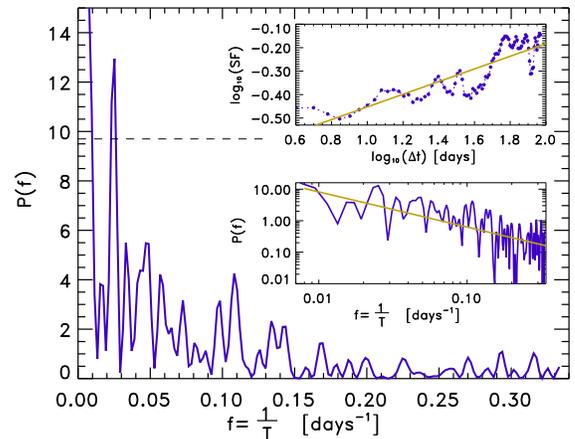}}}%
\vspace{-0.2cm} \caption{
Time series analysis of the LAT light curve presented in Fig.\ref{fig:LATlightcurve}: the periodogram, first order structure function (upper inset) and power spectrum (lower inset). These functions show a variability with a power spectrum consistent with $1/f^{1.3}$ fluctuations. This indicates a variability mode placed between flickering and shot noise. The horizontal dashed line represents the 0.01 false alarm probability threshold (99\% significance that the signal detection is not wrong).
} \label{fig:LATlightcurveTSA}
\vskip -0.1cm
\end{figure}
%
\par A consistent level of variability, with a couple of minor but relevant rapid flares, occurred after the major outburst with fluctuations on timescales of weeks. Renewed activity and increased average brightness from the end of Nov.~2008 were observed. Two rapid flares approached a maximum peak flux of $F_{E>100~\mathrm{MeV}} \sim 2$ \latflux (daily bin estimations), on Sept. 06 (DOY 250),  where a simultaneous cross-correlated optical flare was observed as well, and on Oct.~30 (DOY 304). Visual inspection of the light curve reported in Fig.\ref{fig:LATlightcurve} suggests a period of higher activity beginning in mid Nov.~2008 (after about DOY 320), and, in general, a series of modulations occurring on about a one-month timescale, with faster fluctuations and rapid flare episodes superposed. The power spectral density (PSD) shows a power-law dependence, $P(f) \propto 1/f^{1.3}$. Similar time scale dependence is exhibited by the first order structure function (Fig. \ref{fig:LATlightcurveTSA}) and by the autocorrelation function. More detailed variability analysis for this and other blazars using a longer dataset will be presented in \citet{6monthvariability}.
%
%
\begin{table}[b!!!]
\vspace{-0.1cm}%
\caption[]{Summary of the unbinned likelihood spectral
fit\\ above 100 MeV}\label{tab:spectralfit} 
\vspace{-0.75cm}%
\begin{center}
\begin{tabular}{ll} %
\multicolumn{2}{l}{   } \\ \hline \hline %
Interval [MJD (DOY)] & Best-fit Model and Parameters \\  \hline \hline %
%
%
All observations       & Log-parabola  \\ %
54682.680 (217.680)              & $\alpha= 1.94 \pm 0.05$    \\ %
54775.580 (310.580)              & $\beta= 0.10 \pm 0.02$    \\ %
                       & F$_{E>100~\mathrm{MeV}} = 6.90\pm 0.34 \times 10^{-7}$ [ph cm$^{-2}$ s$^{-1}$]   \\%
\hline 
Outburst/high state        &  Power Law  \\ %
54683.955 (218.955)        & $\Gamma= -2.06 \pm 0.017$    \\ %
54688.985 (223.985)         & F$_{E>100~\mathrm{MeV}}=29.1\pm 1.4 \times 10^{-7}$ [ph cm$^{-2}$ s$^{-1}$]  \\%
\hline 
Post-flare/lower state       &  Log-parabola  \\ %
54689.063 (224.063)               & $ \alpha = 1.87 \pm 0.20$    \\ %
54775.580 (310.580)               & $ \beta = 0.18 \pm 0.08$    \\ %
                     & F$_{E>100~\mathrm{MeV}} = 5.32\pm 1.03 \times 10^{-7}$ [ph cm$^{-2}$ s$^{-1}$]  \\%
%
\hline  \hline
\end{tabular}
\end{center}
\end{table}
%
\subsection{Gamma-ray spectra}\label{subsect:gamma-spectra}
%
%
We have analyzed the time-averaged spectra of PKS 1502+106 for three intervals:  the high-state of the outburst (DOY, 218.95-224.0, i.e. Aug.~05 - Aug.~11, 2008, about 5 days); the post-flare period characterized by an intermediate brightness level and during which simultaneous
\textit{Swift} observations were performed (DOY 224.0-235.42, i.e.Aug.~11-Aug.~22, 2008, about 11.4 days); the longer and heterogeneous period that includes the outburst and the following 4 months of \textit{Fermi}-LAT monitoring (Aug.02 to Dec.07, 2008, $\sim 126$ days, where the source displayed different stages of activity and significant variability). Events below 200 MeV were excluded from these analyses because of calibration uncertainties at those energies. An isotropic background model was used as PKS 1502+106 was very bright relative to other point sources during the period stated above, because is located at high Galactic latitude, and because checks with more complex models provided no significant difference. Furthermore, no appreciable differences were observed using different acceptance cone radii for the event extraction.
\par The spectra for the post-flare and cumulative 4-month datasets
can be consistent with a log-parabola (LP) model, $ dN/dE \propto E^{-(\alpha + \beta \log(E))} $ \citep[see, e.g.][]{landau86,inoue96,fossati00,massaro04,perlman05}. The likelihood ratio test \citep{mattox96} rejects the hypothesis that the spectrum is a power law (null hypothesis) against the one that the spectrum is curved as a log-parabola model. This model is preferred over a simple power-law model at the 11.4 sigma significance level. Broken power-law (BPL) fits show a similar improvement over single power-law
models, but we find no evidence to prefer BPL over the
log-parabola representation. For the full time interval characterized by very high statistics, the logarithm of the likelihood increases significantly when allowing $\beta$ to vary, and an increase of the value for the BPL with respect
to simple power-law model of the same order of the increase for the LP vs. power-law test is not observed. The LP description introduces
the advantage of only one extra parameter (while BPL model adds two parameters) with respect
to the simple power-law model, it allows modeling of mild spectral
curvatures with no abrupt cutoffs, and provides a better phenomenological physical interpretation. On the other hand there can be still caveats
when using \texttt{gtlike} with a broken power-law
model, in particular in determining the break energy when statistics do not allow a high number of energy bins. It is also plausible that an energy spectrum averaged over a long period of time, and containing different activity stages with time varying hardness, may exhibit an apparent curvature. Finally this does not exclude BPL model if the spectrum is extracted in different time intervals.
\par The average spectrum during the outburst state is consistent with a simple power-law model, $dN/dE \propto E^{\Gamma}$. The outburst state shows a rather hard spectrum, suggesting a maximum peak in the MeV energy bands (in the $\nu F_{\nu}$ representation), in agreement with the LP peaks found for the spectra cited above. The extrapolated and averaged fluxes integrated above 100 MeV, and the spectral fit parameters for all three periods are shown in Table \ref{tab:spectralfit}.
%
%
\section{Simultaneous multifrequency observations}\label{sect:simultaneousmwobs}
%
%
%
Because of the reasonably uniform exposure and high sensitivity of the LAT, and the broad-band radio-to-gamma-ray emission of this kind of AGN, simultaneous multifrequency data are very important to the investigation of the physical properties of supermassive black holes and relativistic jets, beyond the benefit of a firm source identification (Section \ref{subsec:association}). With this in mind, several campaigns on a few selected objects, or ToO list of candidates for flaring sources, were prepared pre-launch by the \textit{Fermi} collaboration \citep{tosti07,thompson07}. PKS 1502+106 was a previously unknown $\gamma$-ray source, with no pre-planned multifrequency campaign. But following the LAT outburst (reported in ATel \#1650), a ToO campaign was initiated on Aug.~07, 2008. This was the first \textit{Fermi} multifrequency campaign that had not been planned pre-launch, and saw triggers for ToO pointings by INTEGRAL and \textit{Swift}, long-term radio flux and structure monitoring, as well as optical observations by ground based facilities.
\par The fast response ToO pointing by INTEGRAL provided 200ks of observations during the period Aug.~09, 01:53 UT - Aug.~11, 15:12 UT, 2008 (revolution 711). However, PKS 1502+106 was not detected (preliminarily) by the imager IBIS on board INTEGRAL. Extrapolating the X-ray flux observed by Swift, the hard-X-ray flux had likely already faded to slightly below the IBIS detection threshold in this epoch \citep[more details will appear in][]{pian08}.
%
%
\begin{figure}[t!!]
\hskip -0.4cm
\resizebox{\hsize}{!}{\rotatebox[]{0}{\includegraphics{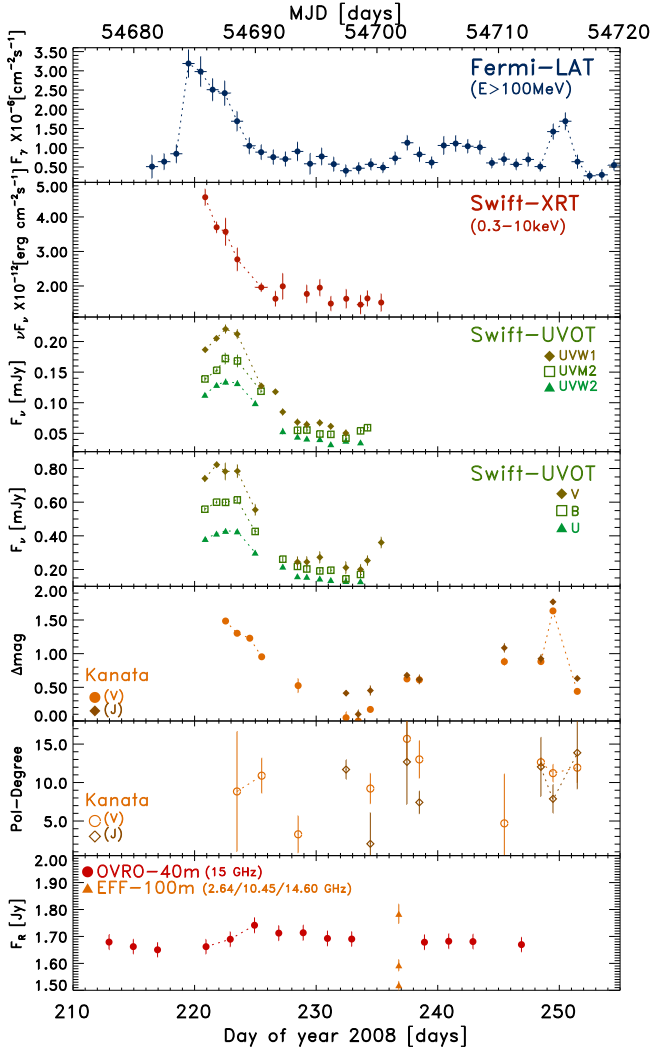}}}
\vskip -0.2cm \caption{Simultaneous gamma-ray and multifrequency
light curves obtained during the multiwavelength campaign of
August 2008 triggered by the high energy outburst discovered by
\textit{Fermi}-LAT. The flux above 100 MeV, the X-ray flux (0.3-10keV) by
\textit{Swift}-XRT, the six-band fluxes monitored by \textit{Swift}-UVOT, the
Kanata-TRISPEC differential relative magnitude light curves (optical $\Delta V$ and near-IR $\Delta J$ bands) and corresponding measures of the linear polarization degree, and the 15 GHz radio
light curve from OVRO 40-m are reported.}
\label{fig:OR103multifreq}
\end{figure}

%
\subsection{Simultaneous X-ray and UV-optical observations and results by \textit{Swift}}\label{sect:swift}
%
%
The \textit{Swift} satellite \citep{gehrels04} performed a
ToO monitoring campaign of PKS 1502+106 with daily snapshots
from Aug. 07 to Aug. 22, 2008. This quite long-term observing campaign by \textit{Swift} allowed extended daily snapshots for about 16 days, using the three instruments onboard: the X-ray telescope (XRT) for the 0.2-10 keV energy band, the Ultraviolet/Optical Telescope (UVOT) for multiband photometry, and the Burst Alert Telescope (BAT) for the 15-150 keV hard X-ray band. BAT
data were not used because of source confusion problems with Mkn 841
which is about a factor of 10 brighter than PKS 1502+106 in the hard X-ray band.
The 16 days of observations by Swift allow for cross-correlation
studies between the $\gamma$-ray, X-ray and UV-optical bands during
both the active flaring stage and the fading post-flare stage of PKS
1502+106.
%

%
\begin{figure}[t!!]
\centering
\resizebox{7.5cm}{!}{\rotatebox[]{-90}{\includegraphics[angle=0]{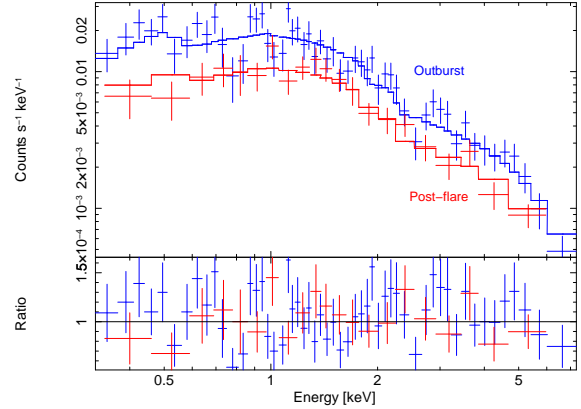}}}\\
\vskip -0.7cm \caption{Swift-XRT combined 0.3-10 KeV spectra of
PKS 1502+106 extracted for the high state (MJD: 54685-54689) and
the subsequent low state (MJD 54690-54701), mapping the X-ray
behavior simultaneous to the LAT flare and to the post-flare, relaxing activity and brightness.} \label{fig:swift-xrt}
\end{figure}
%
%
%
\begin{table}[t!!!]
\vspace{-0.1cm}%
\caption[]{Analysis summary of the simultaneous data obtained by
the XRT instrument on board \textit{Swift}.}\label{tab:swiftXRT2008} 
\vspace{-1.1cm}%
\begin{center}
\begin{tabular}{ll}
\multicolumn{2}{l}{   } \\ \hline \hline %
Obs. id. (date) & Best-fit Model and Parameters \\  \hline \hline %
All observations       & Power Law \\ %
(MJD 54685-54701)      & $\Gamma_{X}=1.53_{-0.07}^{+0.06}$  \\ %
$t_{exp}$:52910s       & $\chi^2_r$=1.05/80 \\ %
                       & F$_{0.3-10~keV}=1.79_{-0.11}^{+0.08} \times 10^{-12}$ \flux \\%
\hline 
Outburst/high state        & Power Law \\ %
(MJD 54685-54689)      & $\Gamma_{X}=1.54\pm 0.08$  \\ %
$t_{exp}$:27680s       & $\chi^2_r$=1.11/52 \\ %
                       & F$_{0.3-10~keV}=2.18_{-0.12}^{+0.15} \times 10^{-12}$ \flux \\%
\hline 
Post-flare/lower state  & Power Law \\ %
(MJD 54690-54701)      & $\Gamma_{X}=1.45_{-0.11}^{+0.12}$  \\ %
$t_{exp}$:25230s       & $\chi^2$=0.76/32 \\ %
                       & F$_{0.3-10~keV}=1.39_{-0.12}^{+0.14} \times 10^{-12}$ \flux \\%
% ...
\hline  \hline
\end{tabular}
\end{center}
\end{table}
%
%

%
\par The XRT was set in photon counting mode, and
the data were processed by the xrtpipeline with the use of
standard software (HEADAS software, v6.4) and standard filtering
and screening criteria. The XRT events in the 0.3--10 keV
energy band were extracted using the {\scshape xrtgrblc} FTOOL
from circular regions centered on the
source position with variable radii depending on the source
intensity and applying correction for vignetting,
Point-Spread Function corrections and hot pixels and columns
with the use of exposure maps. The XRT X-ray flux light curve is shown in the
second panel of Fig. \ref{fig:OR103multifreq}. The 0.3-10keV count rate
of PKS 1502+106 measured by XRT was at a
level 0.05 counts/sec (from our data), up from a level of 0.02
counts/sec (archival past observations). The 0.3-10 keV XRT 16-day
long light curve obtained in August 2008
(Fig.\ref{fig:OR103multifreq}) shows an initial count rate of 0.05
counts/sec, and a gradual decay down to the level of about 0.02
counts/sec.
\par The X-ray spectrum of each observation segment was fitted with an
absorbed power law. Because of the low number of events from the
source, event were not grouped and C-statistics was used,
fixing the column density $N_{H}$ to the Galactic value $N_{HI}=2.19 \times
10^{20}$ cm$^{-2}$ in that direction \citep[in agreement with values used, for example, in ][]{george94,akiyama03}, and using $z=1.839$. The error on the photon
index and the flux (0.3-10keV) is large due to the low statistics. The
background photons were selected in a circular region close to the source.
No significant photon index variation was observed between the high and the low state, while the count rate and flux did vary.
\par The \textit{Swift} Ultraviolet/Optical Telescope (UVOT) photometry
was done using the publicly available UVOT FTOOLS data reduction
suite, and is in the UVOT photometric system described in
\citet{poole08}. The photometric data points were corrected for
Galactic extinction using the dust maps of \citet{schlegel98} and
the Milky Way extinction curve of \citet{pei92}. These
simultaneous multi-band optical and UV data show an increase of
about 2 magnitudes in all filters when compared with the
past-years archival values (i.e., from about 19 to 17 in B band).
The flux light curves in the six UVOT bands are shown in the third and fourth panel of Fig. \ref{fig:OR103multifreq}. These fluxes appears to be
well correlated. A slight rise in flux of 3 days is observed in all the UVOT bands, followed by a fading similar to the flux decrease seen in $\gamma$-rays and X-ray bands. If the time of the observed UV and optical maximum is related to the flare activity at higher energies, this would imply an interesting time lag of about 4 days.
%
%
%
%

\subsection{Simultaneous near-infrared and optical monitoring}\label{kanataresults}
%
%

PKS 1502+106 was also monitored in the optical V and near-infrared J
bands with some photometric and polarimetric snapshots by the
TRISPEC istrument attached to the 1.5-m ``KANATA'' telescope at
the Higashi-Hiroshima Observatory, Japan \citep[][]{watanabe05,uemura08}, within a twofold program of optical follow up for LAT flaring sources and regular monitoring of about 20 blazars.
Imaging relative photometry was performed using some comparison stars in the same field, but
due to the absence of an accurate calibration for this field we prefer to report only the relative magnitude difference $\Delta$mag with respect to the minimum level (fifth panel Fig. \ref{fig:OR103multifreq}).
The optical and near-infrared band imaging photometry is
performed simultaneously in TRISPEC with a unit of polarimetric sequence
(consisting of successive exposures at four position angles of the
half-wave plate, where a set of linear polarization parameters,
Q/I, U/I, are derived).
\par These flux observations,
performed for a longer interval with respect to the \textit{Fermi}-\textit{Swift}
campaign (i.e. until Sept. 22, 2008), show a high correlation between the V- and J-band light
curves and show an optical decay phase comparable to that observed in the UVOT photometric observations (fifth panel of Fig.\ref{fig:OR103multifreq}).
Remarkably, a strong correlation with the LAT gamma-ray light curve is
found, including the first (Sept. 04-07, 2008) of the possible minor flares occurring after the initial large outburst. The observation of a flare in the optical (V) and near-IR (J) bands,
simultaneous with a second $\gamma$-ray flare have the twofold
advantage of providing a validation of such a minor LAT flare as a
real feature displayed by this blazar, and, even more crucially,
in confirming the firm identification of the new gamma-ray source
seen by \textit{Fermi} with the blazar PKS 1502+106.
\par Comparing the Kanata-TRISPEC V-band and J-band colors, the
$V-J$ color index varies between 2.05 and 1.69 (during the Sept.04-07 minor, rapid flare cited above). On the other hand, the degree $P$ of linear optical (in $V,J$ bands) polarization observed (sixth panel of Fig.\ref{fig:OR103multifreq}) , remains rather scattered by error dispersion irrespective the flux level, even during the minor flare mentioned (the maximum degrees recorded during the monitoring were $P(V)_{max}=15\pm3 \%$ and $P(J)_{max}=13\pm 4 \%$).
%
%

\subsection{Simultaneous radio flux-structure data by single-dish
and VLBI observations}\label{sect:radio}
\begin{figure}[b!!]
\hskip -0.1cm 
\centering 
\hskip -0.3cm 
\resizebox{\hsize}{!}{\rotatebox[]{0}{\includegraphics{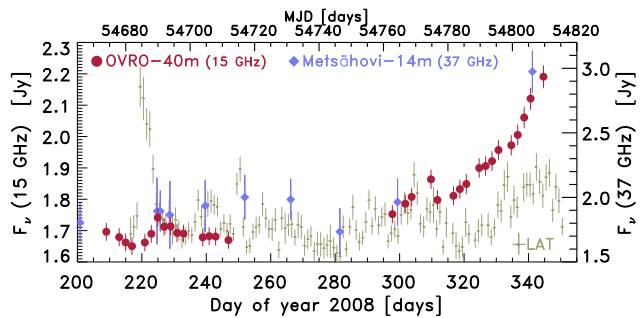}}} 
\vskip -0.1cm 
\caption{Long term radio flux light curve at 15 GHz obtained by the Owens Valley Radio Observatory
(OVRO) 40m dish radio telescope (filled circles), showing the rising part of a
radio outburst started in late November 2008, i.e. almost 4 months
after the gamma-ray outburst detected by \textit{Fermi}. The fill diamonds represent the flux measurements performed by the Mets\"{a}hovi 14-m radio telescope at 37 GHz (right y-axis scale), confirming the start of a radio outburst at a higher frequency. The scaled LAT daily light curve on the same period is reported for comparison (small light grey bars).}\label{fig:radioOVRO}
\end{figure}

%
%
As part of an ongoing blazar monitoring program, the Owens Valley
Radio Observatory (OVRO) 40-m radio telescope has observed PKS~1502+106
at 15~GHz approximately every two days since mid-2007. Flux
densities for the periods from July~26 to September~3, 2008 (MJD
54673--54711) and October~23 to December~9, 2008
(MJD~54762--54809) are shown in Figure~\ref{fig:radioOVRO}.  Flux
densities were measured using azimuth double switching as
described in~\cite{readhead89} after peaking up on-source. The
relative flux density uncertainty for this source is dominated by
a conservative 1.6\% systematic error with a typical
thermal error contribution of 5~mJy. Absolute flux density is
calibrated to about 5\% using the \citet[][]{baars77} model for 3C~286. This absolute uncertainty is not included in the
plotted errors.
\par The measured flux densities in the MJD~54673--54711 (DoY 208-246) time period
fit a 1.69~Jy constant-flux model with $\chi^2/(N-1)=0.70$ ($N=15$
data points). This indicates no statistically significant
variability in this time period.  The beginning of a bright radio
flare is apparent in the MJD~54762--54809 (DoY 297-344) time period with an
increase of at least 30\% over the earlier mean flux density.
%
%
%
\begin{figure}[t!!]
\hskip -0.1cm 
\centering 
\resizebox{8cm}{!}{\rotatebox[]{0}{\includegraphics{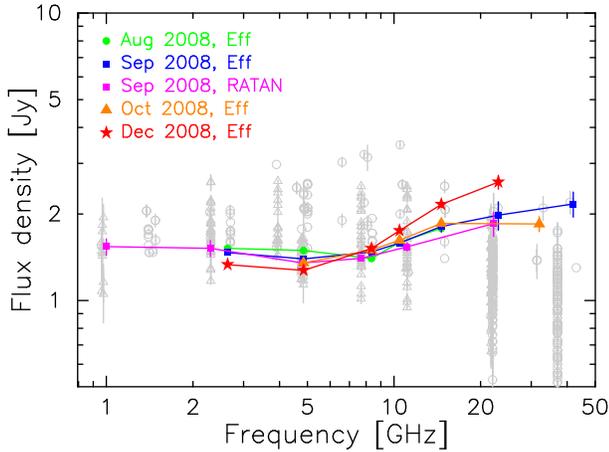}}}\\ %
\hskip -0.3cm 
\vskip -0.1cm 
\caption{Variable broad-band radio spectra
observed with the Effelsberg 100-m and RATAN-600 radio telescopes
simultaneous to the LAT data. Historical RATAN-600 data (grey open
triangles) and archival data from the literature until March 2008 (grey open circles) are shown in the background for comparison.}\label{fig:radiospectra}
\end{figure}
\par A less intensive monitoring at 37 GHz was carried out  with the 13.7m radio telescope at Mets\"{a}hovi Radio Observatory, Helsinki University of Technology, Finland. The flux density scale is set by observations of DR 21, and sources 3C 84 and 3C 274 are used as secondary calibrators. A detailed description on the data reduction and analysis can be found in \citet{terasranta98}.
\par The PKS 1502+106 flare was also followed-up by the Effelsberg 100-m radio
telescope with four multi-frequency radio spectra obtained on
August~23, September~16, October~18, and December~6, 2008
\citep[within the F-GAMMA project, see ][]{fuhrmann07}. Each radio
spectrum consists of simultaneous measurements at various
frequencies between 2.6 and 42\,GHz. The observations were
performed using cross-scans in azimuth/elevation with the number
of sub-scans matching the source's brightness at the given
frequencies. The data reduction was done using standard procedures
described in \citet[][]{fuhrmann08,angelakis08}.

%
\begin{figure}[t!!]
\centering \vskip 0.2cm %
\resizebox{7.5cm}{!}{\includegraphics{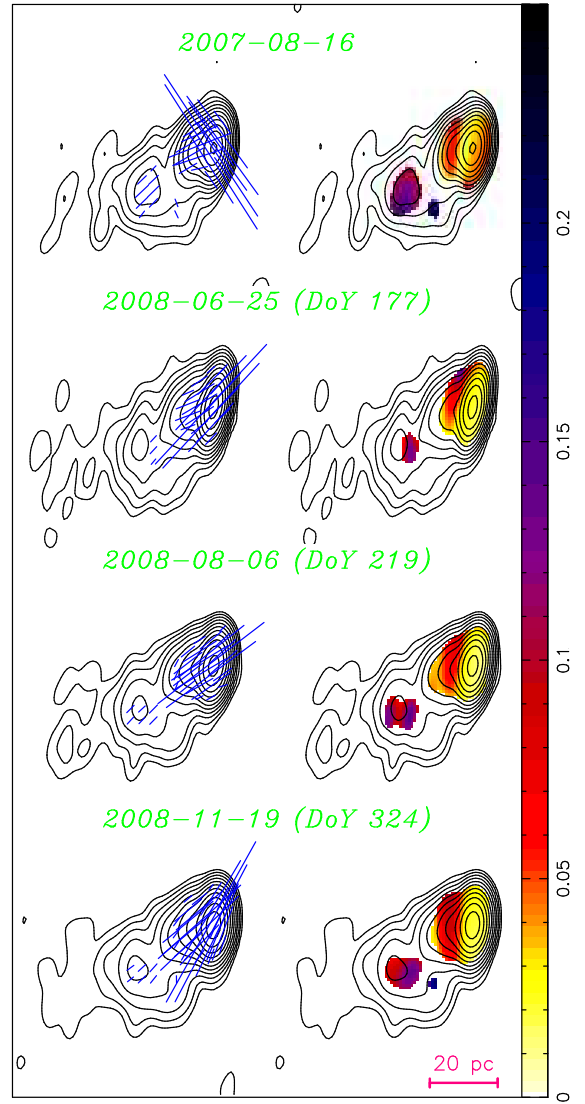}}
\caption{Total intensity and linear polarization images observed
by the VLBA at 15 GHz as part of the large \textit{Fermi}-supporting MOJAVE
program. Naturally-weighted total intensity images are shown by
black contours. The contours are in successive powers of two times
the base contour level of 1.0~mJy\,beam$^{-1}$. Electric
polarization vector directions are indicated on the left hand side
by blue sticks, with their length being proportional to the polarized
intensity. Linear fractional polarization is shown on the right
hand side overlaid according to the color wedge.}
\label{fig:mojave}
\end{figure}
%

%
%
\par Other radio observations are available at six
frequencies between 1 and 21.7~GHz thanks to the 600-m ring
transit radio telescope of the Russian Academy of Sciences
RATAN-600 \citep{korolkov79}, which observed the source on
September~10, 26, and October~2, 2008. A weighted average of these
three observations, is presented in Fig \ref{fig:radiospectra}.
Previous RATAN-600 data which
cover the period between 1997 and March~2008 are also shown for
comparison. The observing methods, the data processing, and the
amplitude calibration is described in \cite{kovalev99}.
\par All single-dish spectra obtained with the Effelsberg 100-m and
RATAN-600 radio telescopes are presented in
Figure~\ref{fig:radiospectra}. Here, no indication of a flare or
strong difference/variability between August and September can be
noted. However, the December~2008 spectrum shows the beginning of
a bright radio flare with a clear spectral steepening towards
higher frequencies ($\nu>10$~GHz). This is in good agreement with
the strong flux density increase seen in the 15 GHz light curve
during November/December (Fig.\ref{fig:radiospectra}).
\par Detailed radio images at sub-milliarcsecond scale of the
PKS~1502+106 superluminal jet were obtained during three epochs in 2008 with \textit{Fermi} already in orbit: on June~25, August~06
(during the maximum peak of the $\gamma$-ray outburst), and
November~19. These observations were performed as part of the
MOJAVE monitoring program conducted with the Very Long Baseline
Array (VLBA) at $\lambda = 2$\,cm \citep{lister05,lister09a} and
provided useful high resolution total intensity and linear
polarization images. These VLBA images close to the $\gamma$-ray
flare are reported and compared to the map obtained one year
earlier in Figure~\ref{fig:mojave}.
The highest integrated flux density value since the beginning of
the 2\,cm VLBA monitoring in 1997 \citep{kellermann04} was
measured on November 19, 2008 (DOY 324) as $F_{15 \mathrm{~GHz}}=2.0$~Jy, with a peak
intensity of 1.6~Jy\,beam$^{-1}$. These values are significantly
higher than the typical level of 1.3~Jy reported in the program
\citep{kovalev05,lister09a,lister09b} and indicate a radio flare happening in
the source VLBI core. The core flux density and brightness
temperature raised to higher values as well which means that the
flare happens in the VLBI core, as expected. These finding are in
good agreement with the single-dish results presented above
(Figure~\ref{fig:radioOVRO} and Figure~\ref{fig:radiospectra}).
The second relevant feature is the direction of the electric
vector position angle (EVPA) in the core region, which rotated
between the 2007 and 2008 epochs by 90~degrees most probably
indicating an opacity change~--- a precursor of an outburst in the
VLBI core.
\par
In summary, our single-dish and VLBI radio monitoring of
PKS~1502+106 simultaneous to the \textit{Fermi} LAT observations has
revealed (i) no significant radio 15\,GHz variability during the
strong LAT $\gamma$-ray flare seen in August 2008, and (ii) a
strong radio flare which becomes clearly visible at $\nu>10$~GHz
during October/November~2008 (Figure~\ref{fig:radioOVRO}) with a rise phase
lasting for at least 20 days. If this flaring behavior is
associated with the bright $\gamma$-ray flare of August~2008, a
delay of more than three months ($\sim 98$ days if the lag between the starting days of the $\gamma$-ray outburst, DoY 218.2, and the 15 GHz outburst, DoY 316.7, is considered) could be explained by opacity
effects in the core region of the source \citep[e.g.,][]{aller99}.
Although we cannot exclude the possibility of radio activity at 15 GHz
during the OVRO 40-m observations outage (September 3 - Otober 23 2008) that might also be associated with the gamma-ray flare of August 2008, OVRO data before and after the outage as well as September and October Effelsberg 100-m data are consistent with very little change in the 15 GHz flux over this time period. The three flux measures at 37 GHz obtained at the Mets\"ahovi radio observatory during the OVRO outage, confirms this trend with very little variability during this period. However, the observed radio flare could also be associated with,
e.g., the more recent, prominent variability seen in the LAT
$\gamma$-ray data during November/December
(Figure~\ref{fig:LATlightcurve}, DoY\,$\sim$\,320--333). A more
detailed analysis of such possible correlations and the source's
overall radio/$\gamma$-ray behavior seen with LAT and simultaneous radio observations over
a longer period of time will be the subject of a subsequent work.
%
%

%
\section{Archival multiwaveband data}\label{sect:otherMW}
%
A full multiwavelength analysis dedicated on PKS 1502+106 is available only
from \citet{george94}, where old archival and broadband radio to X-ray data obtained in 1993-94 were presented.
Data and analysis on PKS 1502+106 reported in other papers are mostly limited to the radio regime. In order to compare our multifrequency findings with the past, and to form a more complete characterization of this blazar, we briefly present here, for the first time, results from unpublished past
observations by INTEGRAL, XMM-Newton, Swift, and Spitzer space telescopes
performed in 2001, 2005 and 2006.

%
\subsection{INTEGRAL observations in 2006}
%
The sky region containing PKS 1502+106 was observed in
2006 by IBIS (83ksec, MJDs 53760.4 to 53762.4, Jan.25-27, 2006),
and a new soft $\gamma$-ray source (IGR J15039+1022) was detected
with a flux density of 1.6 mCrab in the 18-60 keV energy range
(corresponding to $1.2 \times 10^{-11}$ erg cm$^{-2}$ s$^{-1}$,
see ATel \#1652). This IBIS source was identified with Mkn 841,
a Seyfert galaxy known to display a well detected high
energy cut-off around 100 keV \citep{petrucci02}, making it
unlikely to emit in the $\gamma$-ray domain. The angular distance of PKS 1502+106
from IGR J15039+1022, $\sim 11'$, points to a clear non-detection
during this Jan. 2006 INTEGRAL observation, while a 2$\sigma$ upper limit for PKS 1502+106 of 0.7 mCrab
in the range 18-60 keV ($0.52 \times 10^{-11}$ erg cm$^{-2}$ s$^{-1}$) is inferred.
%
\subsection{XMM-Newton and Swift}
%
Four serendipitous, unpublished XMM-Newton X-ray observations of PKS
1502+106 by the EPIC (MOS detector only) camera are available as the source was in the frame of the target Seyfert galaxy Mkn 841. PKS 1502+106 was always on the border of the MOS chips and out of the PN frame, and therefore subject to low X-ray statistics, regardless of its intrinsic brightness.  The four X-ray EPIC-MOS
observations (three in 2001 and one performed in Jul.17, 2005, see
Table \ref{tab:xmm} for an analysis summary) do not show variations in the  0.2-10 keV photon index, while the 0.2-10 keV flux intensity varied by a factor of a few (in the range $3.5-6.8 \times 10^{-13}$ \flux in 2001, comparable to the lower states observed by ASCA, and a mildly active state with $1.1 \times 10^{-12}$ \flux in July 2005, Table \ref{tab:xmm}).
%
\begin{table}[b!!!]
\vspace{-0.1cm}%
\caption[]{Analysis summary of the EPIC-MOS instrument
observations (Jan. 2001 and Jul. 2005) on board of XMM-Newton.}\label{tab:xmm} 
\vspace{-1.1cm}%
\begin{center}
\begin{tabular}{ll}
\multicolumn{2}{l}{   } \\ \hline \hline %
Obs. id. (date) & Best-fit Model and Parameters \\  \hline \hline %
%
ObsID 0070740101            & Power Law \\ %
(Jan 13, 2001, 09:20 UTC)   & $\Gamma_{X}$= 1.6 $\pm$ 0.2 \\ %
                            & $\chi^2_r$=1.69/11 \\ %
                            & F$_{0.2-10~keV}=3.5 \times 10^{-13}$ \flux \\%
\hline
ObsID 0070740301           & Power Law \\ %
(Jan 14, 2001, 00:30 UTC)   & $\Gamma_{X}$= 1.7 $\pm$ 0.2 \\ %
                           & $\chi^2_r$=1.27/17 \\ %
                           & F$_{0.2-10~keV}=6.8 \times 10^{-13}$ \flux \\%
\hline
ObsID 0112910201            & Power Law \\ %
(Jan 13, 2001, 04:58 UTC)   & $\Gamma_{X}$= 1.6 $\pm$ 0.2 \\ %
                            & $\chi^2_r$=1.05/8 \\ %
                            & F$_{0.2-10~keV}=3.6 \times 10^{-13}$ \flux \\%
\hline
ObsID 0205340401            & Power Law \\ %
(Jul 17, 2005, 06:32 UTC)   & $\Gamma_{X}$= 1.69 $\pm$ 0.08 \\ %
                            & $\chi^2_r$=0.99/58 \\ %
                            & F$_{0.2-10~keV}=11.0 \times 10^{-13}$ \flux \\%
\hline  \hline
\end{tabular}
\end{center}
\end{table}
%
%
\par PKS 1502+106 was also observed twice in the past with the \textit{Swift}-XRT
as a fill-in target (TargetID: 36388), showing a fainter X-ray
flux (0.02 counts/sec) than the flux recorded in the August 2008 campaign observations (0.05 counts/sec).

%
\subsection{Spitzer observations and the multifrequency behavior on Jul-Aug.2005}\label{subsect:spitzer_and_2005_behavior}
\begin{figure}[t!!]
\hskip -0.1cm
\resizebox{\hsize}{!}{\rotatebox[]{0}{\includegraphics{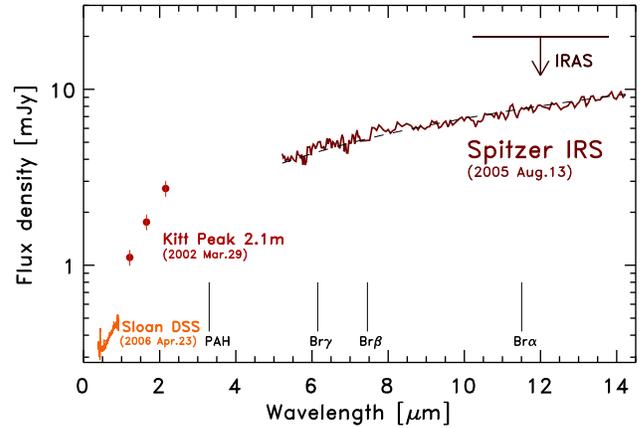}}}
\vskip -0.1cm \caption{The unique observed mid-IR spectrum in the range 5-14 $\mu$m
obtained by the Spitzer Infrared Spectrograph (IRS) low resolution
$(R=60-130)$ module, in August 13, 2005. The position of redshifted Brackett emission lines and PAH line are indicated, even if they are not detected in the IRS spectrum that is consistent with a simple power law model. In addition the optical spectrum by the SDSS on April 23, 2006, and high precision JHK$_{\mathrm{S}}$ photometric flux measurements of the Kitt Peak National Observatory (KPNO) 2.1m telescope of March 29, 2002 are also reported.} \label{fig:spitzerIRS} \vskip -0.1cm
\end{figure}

%
In the past, only upper limits  in the far-/near-infrared bands by IRAS
were available for PKS 1502+106 \citep[][and Figures \ref{fig:spitzerIRS} and \ref{fig:SEDLAT2008}]{neugebauer86}.
PKS 1502+106 was observed serendipitously in the mid-infrared band for
the first and only time by Spitzer on August 13, 2005, 09:10-09:18
UT (PID 117, AOR Request Key 5011456). The Infrared
Spectrograph \citep[IRS,][]{houck04}, low resolution (R=60-130)
module, recorded the mid-IR spectrum from 5-14 $\mu$m (shown in
Fig.\ref{fig:spitzerIRS} with the optical SDSS spectrum and near-IR photometric data point). High resolution IRS module spectra were also taken,
but could not be used since there were no
accompanying background observations. The Short-Low (SL) coadded
2D spectra were reduced using the standard Spitzer IRS pipeline
(ver. S17.2).
%
%
Background was subtracted using the two nod positions along the
slit. The spectra were extracted and flux-calibrated with SPICE
ver.~2.1.2, in a standard, expanding point-source aperture. The
two spectral orders match well at 7 $\mu$m, indicating a
well-pointed observation. The mid-IR continuum of PKS 1502+106
rises to the near infrared and appears to be rather featureless,
consistent with pure synchrotron emission (a power law
$F_{\nu}\propto \nu^{-0.9}$ in the 5-14 $\mu$m range, Fig.\ref{fig:spitzerIRS}). The
wave-like deviation of the data can be simply explained as
wavelength-dependent spectrograph slit losses, while the
red-shifted Brackett emission line series (like the 3.3 $\mu$m PAH
feature) falling in this wavelength range are indicated, but they
are not well detected in the spectrum. This
IRS spectrum is similar to other blazars which have been
observed by Spitzer, including BL Lac and 3C 454.3
\citep{leipski08,ogle08}, while the near IR (J,H,K) flux data
from the Kitt Peak National Observatory (KPNO) 2.1m telescope
\citet{watanabe04} reported in the same figure indicate a lower
flux state and steeper H-K$_{\mathrm{S}}$ (1.65-2.15~$\mu$m), spectral index $\left(\alpha_{(H-K_{\mathrm{S}})}=1.66\right)$.
\par The July-August 2005 SED assembled with these XMM-Newton (July 17, 2005) and Spitzer (August 13, 2005) observations (joined with a couple of radio-optical observations in these two months by the Mets\"ahovi, RATAN and Catalina observatories) is consistent with a low or mildly active
stage that can be explained by a simple SSC model (inset plot in Fig.\ref{fig:SEDLAT2008}). Based on these data there
are therefore no hints for a deviation from a SSC scenario in this
flat spectrum radio quasar. In any case, this SED cannot provide
significant constraints, because it contains observations obtained over
about two months and does not include any $\gamma$-ray data, aside
from the older (1992) EGRET upper limit \citep[calculated with the prescriptions in ][]{thompson96}.

\par XMM-Newton observations pointed out an (0.2-10 keV) photon index that
shown almost no variations, with values ($\Gamma_{X} \sim 1.7$) in
agreement with the previous X-ray observations performed by ROSAT
and ASCA \citep{george94,watanabe04}. The integrated X-ray flux
F$_{0.2-10~keV}=1.0 \times 10^{-12}$ suggests a mildly active
state during this observation, comparable to the \textit{Swift} XRT
spectrum observed in the days soon after the LAT outburst (red
filled symbols, data points on Fig.\ref{fig:SEDLAT2008}). More complicated
and possibly more accurate emission models, beyond the
SSC, can be therefore investigated for the first time only thanks
to the \textit{Fermi} multifrequency campaign of August 2008, whose SED is
described in the next section. The \textit{Swift} XRT simultaneous spectrum of Aug.2008
has a slightly harder spectrum with respect to these archival XMM-Newton observations
while the XRT flux was about one order of magnitude
greater than the flux observed in the past by XMM-Newton, ASCA and
ROSAT.

%
%
\section{Discussion}
%
\subsection{Gamma-ray outburst and longer-term variability}\label{subsect:gamma-variability}
%
%
During the first several months of the LAT survey, PKS 1502+106 was one of the brightest, as well as the most luminous, blazar in the MeV--GeV band. The threefold flux increase in $\lesssim 12$ hours  between Aug. 05 and Aug. 06, 2008 (DoY 218-219, Fig. \ref{fig:LATlightcurve}) constrains the rest-frame size ($R^\prime$) of the flaring region: $R'\leq c \Delta t \mathcal{D}/(1 + z) \simeq 6.8 \times 10^{15} $ cm (where
$\mathcal{D}= 1/({\Gamma(1-\beta \cos \theta)})$ is the macroscopic Doppler factor, $\Gamma$ the bulk Lorentz factor and $\theta \simeq 1/\Gamma$ the angle of sight).
The value $\mathcal{D} \simeq \Gamma \simeq \beta_{app} =14.8 $ is assumed from MOJAVE VLBA measurements \citep{lister09b}, where $\beta_{app}$ is the kinematic apparent jet speed in units of c. In this high speed regime an upper limit on the viewing angle can be also estimated: $\theta < 2 \arctan (1/\beta_{app}) = 7^{\circ}.7 $. This is consistent with an independent estimation of the Doppler factor based on the 37 GHz flux variability, made in \citet{hovatta09}: a factor $\mathcal{D}_{var}=12$ and $\beta_{app\_var} =14.6$, with brightness temperature $T_{br}= 8.7 \times 10^{13}$ $^{\circ}$K, and angle of sight $\theta=4.7^{\circ}$.
\par Assuming the ``concordance'' cosmology (Sect. \ref{sect:introduction} and $z=1.839$), the luminosity distance of PKS 1502+106 is $d_{L}=14.2$ Gpc, and the inferred, apparent and isotropic, monochromatic luminosity at $E_{0}=100$ MeV of PKS 1502+106 during the outburst phase is $L_{E>100} \simeq 4\pi d_{L}^{2} \cdot \left( \Gamma -1 \right) E_{0}F_{E>E_{0}} \simeq 1.1 \times 10^{49} $ erg s$^{-1}$,, where the average flux $F_{E>E_{0}} =  2.91 \times 10^{-6}$ ph cm$^{-2}$ s$^{-1}$ in the outburst interval (DoY 2008: 218.95 - 224.0) is used. The bolometric luminosity is expected to be even higher than this value, since the measured LAT spectrum appears to be beyond the peak of the high energy component, and therefore this LAT blazar has probably one of the highest $L_{\gamma}/\Delta t $ ratios ($2.5 \times 10^{43}$  erg s$^{-2}$) known in the MeV-GeV regime.
\par Relativistic motion provides a solution for the problem of intrinsic excess absorption by pair-production in powerful $\gamma$-ray sources like PKS 1502+106 \citep[see, e.g. ][]{mattox93,madejki96} which have a significant $L_{BLR}$.
Adopting the flux tripling time scale of the outburst rise (i.e. $\Delta t = 12$ hours), and the outburst state averaged X-ray flux (F$_{0.3-10~keV}=2.18_{-0.12}^{+0.15} \times 10^{-12}$,  Table \ref{tab:swiftXRT2008}) at the observed photon frequency $\nu_{X}=10^{18}$ Hz, the minimum Doppler factor $\mathcal{D}$ required for the photon-photon annihilation optical depth to be $\tau_{\gamma \gamma} \leq 1$ can be estimated. Using the derived relation $1=\tau_{\gamma \gamma} \simeq  \sigma_{T} d^2_L F_X  / \left( 3 \Delta t ~c^2 E_X \mathcal{D}^4 \right)$ and taking the region size $R= c\Delta t \mathcal{D} /(1+z)$, the source-frame photon energy $E'_{X}=(1+z) h \nu_X/ \mathcal{D}$ and the intrinsic X-ray luminosity $L'_X=4\pi d^2_L \mathcal{D}^{-4} F_X$ we obtain $\mathcal{D} \gtrsim 7.7$ (omitting the requirement of
co-spatiality of the X-ray and $\gamma$-ray emission regions relaxes this limit). This is in agreement with the values found from radio flux-structure observations and with the SED modeling parameters found.
\par The asymmetry of the August 5--6 $\gamma$-ray outburst can suggest a more complex emission geometry than a simple one-zone model. The temporal structure --- $\sim 0.5$ day rise, followed by a $\sim 4.5$ day decay where a $\sim 2.5$ day intermediate-level plateau is likely observed --- implies particle acceleration and cooling times that are greater than the light crossing time, i.e., $t_{\rm inj}, t_{\rm cool} > R/c$, where these quantities are evaluated in the jet comoving frame.
A synchrotron self-Compton (SSC) emitting blob in the jet should be relatively confined ($<0.01$ pc), although relativistic beaming would permit the region to be as much as an order of magnitude larger. The hinted intermediate plateau could mark a twofold active region, and two SSC emitting components. Descriptions making use of a multi-zone SSC or multi-emission component SSC (second order, superquadratic components) are reported, for example, in \citet[][]{sokolov05,georganopoulos06,graff08}.
On the other hand, if the injection of relativistic electrons is impulsive and repeated, some single-zone SSC models already predict plateaus during an outburst \citep[see e.g., ][]{chiaberge99,boettcher02,sokolov04}. At lower frequencies (IR-optical-UV), where cooling times are longer, the electron distributions corresponding to different injections can build up, and the memory of the individual injection phases can be lost, providing a smoother decay without intermediate plateaus (as shown in X-ray, optical-UV light curves by \textit{Swift}, Fig. \ref{fig:OR103multifreq}). Apparent delays (like the 4-day lag hinted by UVOT data) can also be explained within this scenario.
\par On the other hand this asymmetric (fast rise, slower decay) shape of the $\gamma$-ray outburst can be also an evidence for a dominant contribution by Comptonization of photons produced outside the jet (Sect. \ref{subsect:gamma-x-uv-correlation} and \ref{subsect:SED}) during this event, as predicted for example in \citet{sikora01}. Gamma-ray flares produced by short-lasting energetic electron injections and at larger jet opening angles are predicted to be more asymmetric showing much faster increase than decay, the latter determined by the light travel time effects.
\par The ``harder when brighter'' trend of the gamma-ray photon index during the outburst (right inset panel of Fig. \ref{fig:LATlightcurve}), hints a narrow hysteresis evolution of the spectral index against the flux, a signature produced by non-thermal cooling and high to low energy propagation of the electron injection rate \citep{kirk98,georganopoulos98,boettcher02}. The photon index extracted with a power-law model over the daily bins, was found quite scattered irrespective of the flux level in the remain part of the light curve following the outburst.
\par The outburst of Aug.~2008 appeared to have ignited an enduring relaxing state of $\gamma$-ray brightness and activity, during the following four months (Sections \ref{subsect:gamma-lightcurve} and \ref{subsect:gamma-variability}). The $1/f^{a}$ (with $a \sim 1.3$) PSD points out a general fluctuation mode placed between a pink-noise (flickering) and a random-walk, (Brownian motion or brown noise), staggered by some rapid flares, similar to the long-term variability of blazars observed in radio and optical wavebands \citep[for example, ][]{hughes92,ciaramella04,terasranta05,ciprini07b,hovatta07}. In contrast, this variability mode is rather different than the full Brownian regime shown by the short-term (intra-hour resolution) light curve of the large TeV outburst of PKS 2155-304 \citep[][]{aharonian07}. This PSD indicates the occurrence frequency of a specific variation is inversely proportional to its strength, as found in processes driven by stochastic relaxation, and rapid flares/outbursts (phenomenologically called intermittence), are not occasional events produced by physical processes of different nature with respect to the mechanism responsible for the long-term flickering, but can be considered as statistical tails of the same dynamic process, possibly connected to disk or jet instabilities, to viscosity and magnetic turbulence, or to inhomogeneities and shocks \citep[for example ][]{begelman84}. Even in the case of quite nonhomogeneous structures, the jet is seen under a small viewing angle ($\theta < 7^{\circ}.7 $ here), therefore flaring events and variability trends are the result of emission components originating from different regions, excluding, in most cases, causality.
\subsection{Gamma-ray and radio connection}\label{subsect:gamma-radio}
%
The absence of significant radio 15\,GHz and 37\,GHz variability before,
during and immediately after the LAT outburst (Fig.~\ref{fig:radioOVRO}), and strictly correlated to the near-IR to $\gamma$-ray outburst, has consequences for opacity at lower radio frequencies. However, the beginning of the strong radio flare seen in the Oct.-Dec. 2008 data of Fig.
\ref{fig:radioOVRO} and Fig. \ref{fig:radiospectra}, could be associated with the period of increased $\gamma$-ray activity seen after $\sim $DoY 330 (Nov.~25, 2008, Fig.~\ref{fig:LATlightcurve}). If the beginning of the radio outburst in Fig. \ref{fig:radioOVRO} is associated with the LAT outburst, this would imply a delay of more than about 98 days (Sect. \ref{sect:radio}), meaning that the radio emission region becomes transparent at the cm and longer wavelengths significantly after the outburst seen at wavelengths where the source is truly optically thin (mm, IR, optical bands).
\par The rotation direction of the electric vector position
angle (EVPA) in the milliarcsecond-scale core of PKS 1502+106 between 2007 and 2008 was a possible precursor signature of an outburst that occurred in the core (Fig.\ref{fig:mojave}). The electric polarization vectors appear well aligned to jet axis in the June 25, 2008 map and even more in the Aug. 06, 2008 map. This latter map represents an unprecedented example of radio-structure snapshot truly simultaneous to the peak of a MeV-GeV outburst (Fig.\ref{fig:mojave}). About 3 months after the outburst, on Nov. 11, 2008, the alignment is again decreasing. These interesting findings are in agreement with the scenario that assumes very bright $\gamma$-ray flares occurring after the ejections of superluminal radio knots, with accompanying increases
in polarized radio flux, and a field ordering and alignment with
respect to the jet axis. A correspondence between variations in polarization direction and intensity in different bands at parsec scales can help localize the primary site of the high energy emission. We found that the prominent $\gamma$-ray outburst and the possible 3-month delayed radio-outburst are likely produced in the core of the parsec-scale jet of PKS 1502+106. More details on connections between VLBI radio structures and high energy emission, and their interpretation are reported recently in \citet{jorstad07,darcangelo07,marscher08,kovalev09}. To test if this is indeed the case for the discussed gamma-ray flare, VLBA monitoring of 1502+106 is continued at 15 and 43 GHz.
%
%

\subsection{Gamma-ray, X-ray and UV-optical cross correlations}\label{subsect:gamma-x-uv-correlation}
%

\begin{figure}[t!!]
\centering 
\hspace{-0.4cm} 
\resizebox{4.5cm}{!}{\rotatebox[]{0}{\includegraphics{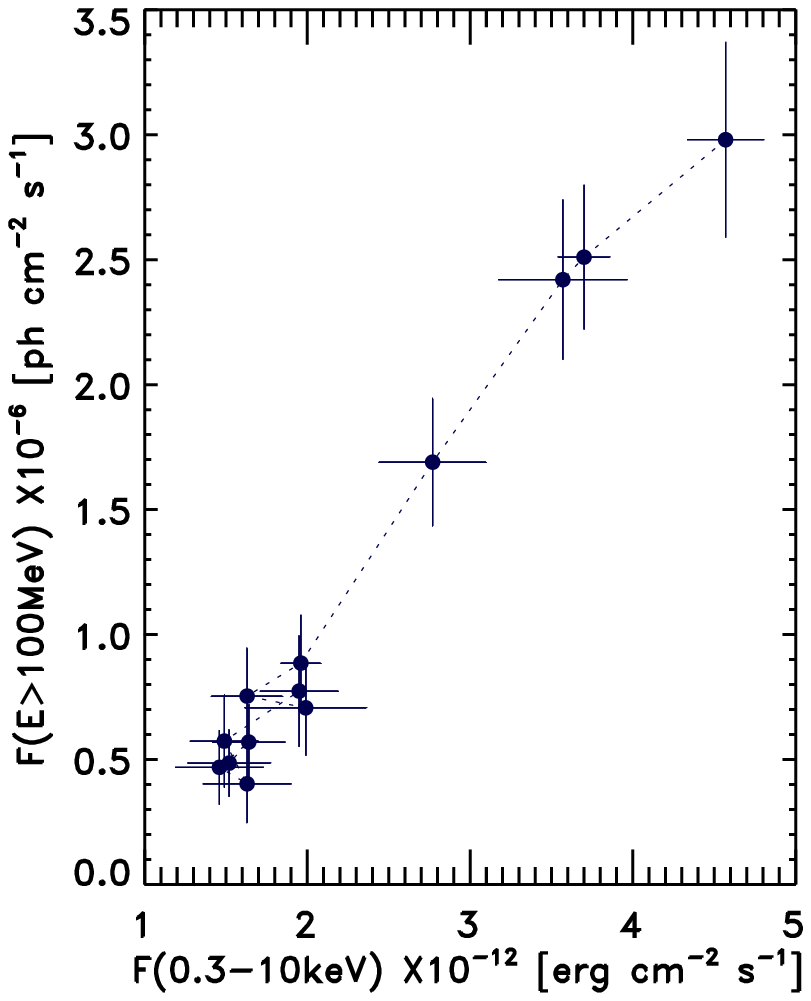}}}%
\hspace{-0.7cm} 
\resizebox{4.5cm}{!}{\rotatebox[]{0}{\includegraphics{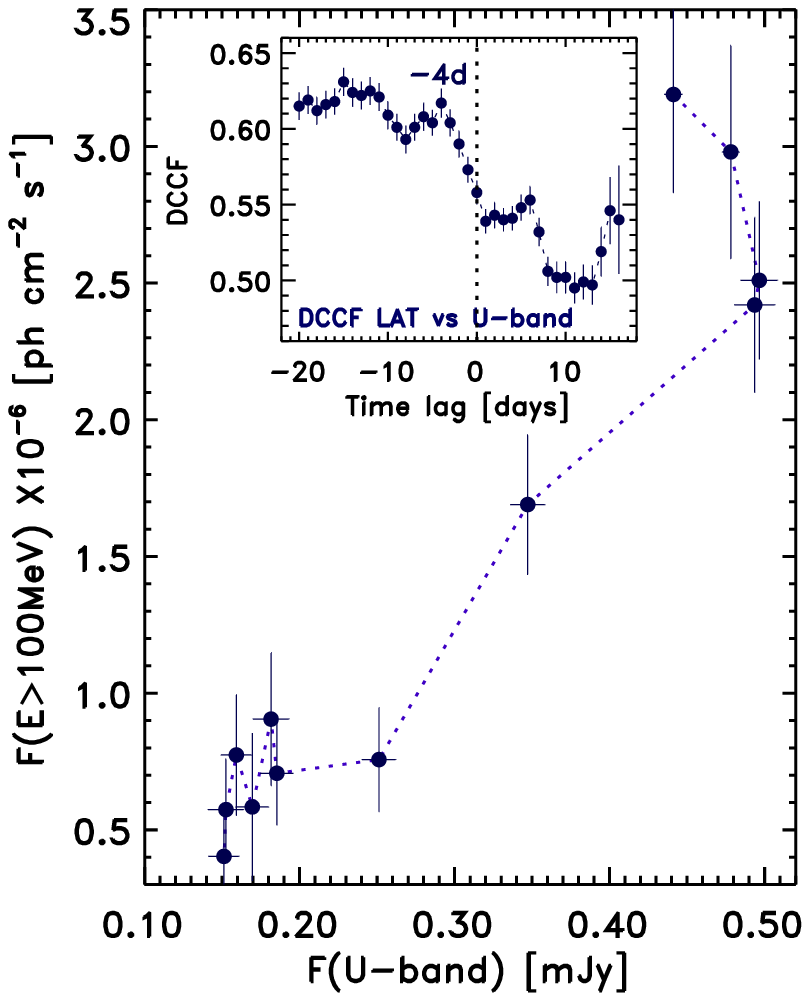}}}
\\%
\vskip -0.2cm \caption{\textit{Left panel:} the gamma-ray flux
measured by \textit{Fermi}
LAT versus the X-ray flux measured by \textit{Swift}-XRT. The
cross-correlation, without lags is well displayed. \textit{Right
panel:} the LAT gamma-ray flux versus the UV flux measured by
\textit{Swift}-UVOT in the representative U-band, and the discrete cross
correlation function diagram (DCCF, inset plot) between the
gamma-ray flux and the U-band flux. Here a clear correlation with a 4 day delay of the UV emission flare with respect to the
gamma-ray flare is suggested.} \label{fig:swif-lat-crosscorr}
\end{figure}

\begin{figure*}[ht!!]
\hskip 0.1cm
 \centering
\resizebox{16cm}{!}{\rotatebox[]{0}{\includegraphics{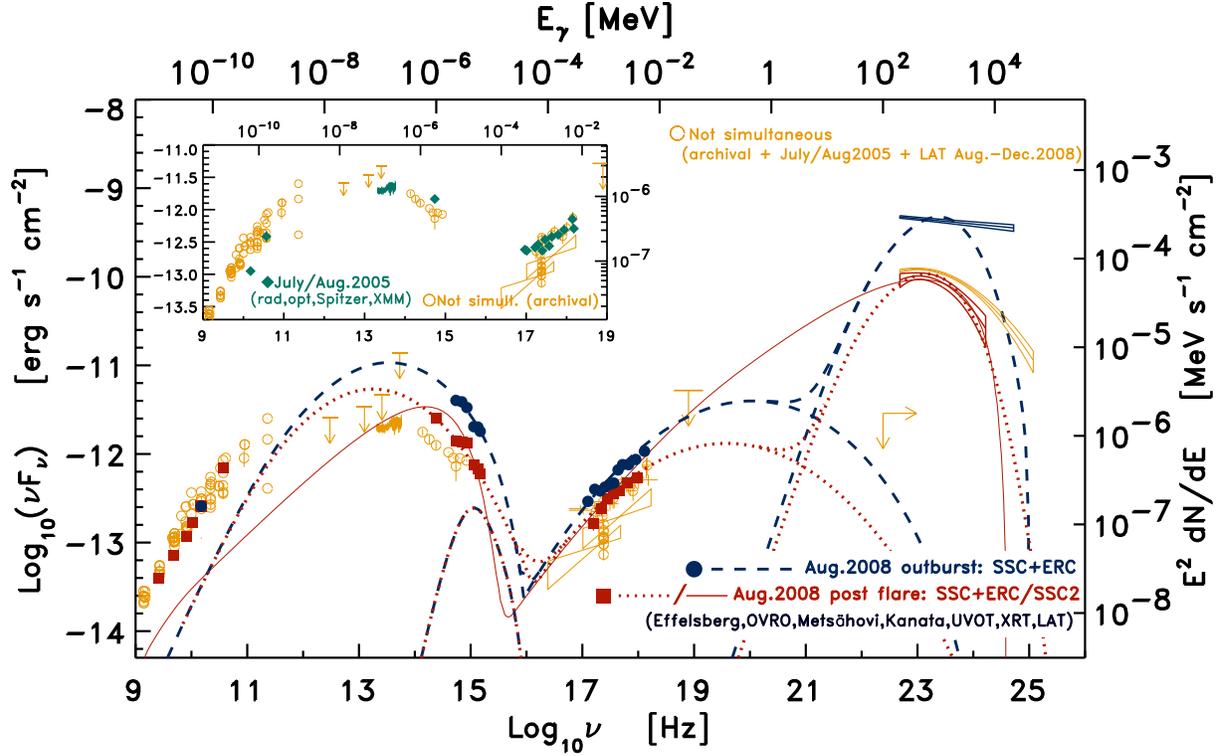}}}\\
\vskip -0.2cm \caption{\textit{Major panel:} Overall radio-to-gamma-ray spectral energy distribution (SED) of PKS 1502+106 assembled with data from the
2-week \textit{Fermi} LAT multifrequency campaign of August 2008. The two
representative, time-averaged, states for this campaign (time intervals outlined in Fig.\ref{fig:LATlightcurve}), i.e. the
high state (Aug.~05-10, 2008, DoY 2008: 218.95 - 224.0;  blue filled circle symbol), and the post-flare (intermediate brightness) state (Aug.~11-22, 2008; DoY 2008: 224.0 - 235.42; red filled square symbol), are represented along with their SSC and ERC model attempts. Archival non-simultaneous data (including
the Jul.-Aug.~2005 data, and the whole 4-month cumulated LAT
spectrum, open orange circle symbols and strip) are reported in the background
for comparison. The high Compton dominance and gamma-ray bolometric luminosity reached during the outburst is evident. The SED of the outburst state is reported with a superposed one-zone ``SSC+ERC'' model fit attempt (where the SSC dominates the radio-to-X-ray SED and the ERC produces the gamma-ray component, blue dashed line), while the SED of the post-flare state is reported with two possible models superposed: the same SSC+ERC modeling (SSC radio-to-X-ray and ERC for gamma-ray band, red dotted line) and also a pure one-zone SCC attempt (labeled with ``SSC2'', a second type of stand-alone SSC model for the entire radio-to-gamma-ray SED, continuous red line).
\textit{Inset panel:} The July-August 2005 non-simultaneous SED
(filled orange circles) assembled with the XMM-Newton EPIC-MOS
data from Jul.~17, 2005, the Spitzer IRS observation from Aug.~13, 2005,
radio flux data from Mets\"ahovi
and RATAN radio observatories, and optical data from the Catalina
Sky Survey. Archival non-simultaneous data (open circles) are reported in background. This two-month averaged SED
is consistent with a low or mildly active stage (suggested by
the X-ray flux above $1 \times 10^{-12}$ \flux) and can still be explained by a pure one-zone SSC model.
} \label{fig:SEDLAT2008}%
\vskip 0.2cm
\end{figure*}

During the EGRET era a similar degree of simultaneous X-ray and $\gamma$-ray monitoring was achieved, for instance, for the blazar 3C 279 \citep{wehrle98,hartman01}, a typical $\gamma$-ray powerful FSRQ resembling PKS 1502+106. The correlation found in PKS 1502+106 between the $\gamma$-ray flux and the
X-ray and UV-optical emission during the 16-days of \emph{Swift} follow up, is evidenced by the multi-panel and multifrequency light curves in Fig.~\ref{fig:OR103multifreq}.
The apparent linear correlation between the emission in the LAT band and the X-rays
(Fig.~\ref{fig:swif-lat-crosscorr}, left panel) suggests that the observed X-ray and
MeV-GeV photons may be part of a single SSC component (continuous line, labeled ``SSC2'', on Fig.\ref{fig:SEDLAT2008}). This could also explain the nearly constant X-ray (0.3-10 keV) spectral index observed during the outburst with respect to the post-flare phase.
\par The $\gamma$-ray peak power observed during the outburst was $\nu F_{\nu} \sim 10^{49}$ erg s$^{-1}$ , and decreased to less than about $2 \times 10^{48}$ erg s$^{-1}$ in the post-flare period, meaning a difference of at least a factor of $\sim 6$ (Fig.~\ref{fig:SEDLAT2008}). The difference between these two emission states in X-ray output was instead a factor of $\sim 2.5$, and a factor of $\sim 3.5$ in the optical-UV band (frequencies above the synchrotron peak). These sub-quadratic differences between the outburst and the subsequent state could be still described by SSC descriptions. The strict correlation between the X-ray flare and the $\gamma$-ray outburst supports, at least, the dominance in the SED (Fig.\ref{fig:SEDLAT2008}) of the SSC, in-jet, emission from radio to X-ray bands in agreement with results detailed in \citet{sikora01}. A simple single-zone SSC model has nevertheless problems explaining and reproducing the large $\gamma$-ray dominance observed during the outburst, and could require very sub-equipartition magnetic fields.
%
%
%
%
\par Fig.~\ref{fig:OR103multifreq} and the right panel of Fig.~\ref{fig:swif-lat-crosscorr} show the flux monitored by UVOT (U-band reported there, but similar results are found in the other 5 filters), indicating the same strong correlation with respect to $\gamma$-rays, but with a possible time lag of about 4 days (also hinted at by the peak on the discrete cross-correlation function, DCCF, inset plot).
This possible time lag can be reasonable only by assuming that the optical-UV brightness during the start and rise of the gamma-ray outburst (i.e. during DoY 216-219) was comparable or lower than the flux observed during the first UVOT observation (performed on DoY 220.8). Relative delays between the synchrotron (our \textit{Swift} UVOT data) and the inverse Compton (our \textit{Swift} XRT and \textit{Fermi} LAT data) counterparts, are dominated by energy stratification and geometry of the emitting region. In SSC-dominated zones the synchrotron emission is co-located with the inverse-Compton production site, and
such 4-day optical-UV to $\gamma$-ray lag would depend on light-travel time effects in the the emitting region (characterized by size $R$ and viewing angle $\theta$), and by the particle cooling time and decay time of the synchrotron and inverse Compton (IC) emissions. When the region is not so compact, both decay times can be comparable to the apparent light crossing time $R/c$, and significant shifts between light curves at these different energy bands are expected. Synchrotron flares (our optical-UV data) can be delayed with respect to the IC $\gamma$-ray flares \citep[e.g.][]{chiaberge99,sokolov04}. Rather long time lags such as this could also be explained by a prolonged disturbance traveling down the jet. The disturbance, being radially inhomogeneous in both density and velocity, could induce shocks and collisions leading to the formation of two adjacent emission zones with similar properties (multi-zone SSC), thus explaining the flare shape asymmetry and the intermediate-level plateau during the decay (Section \ref{subsect:gamma-variability}).
\par Unlike SSC emission, outbursts dominated by inverse Compton external radiation involve a constant field of seed photons, and are not delayed by light-travel time of photons. The resulting frequency stratification behind the shock front could extinguish an outburst first at the highest energies, then progressing to lower frequencies as time advances \citep{sokolov05}, and energy-dependent radiative losses induce delays in the declining part of the emission produced by the lower energy particles. The variability patterns shown by our data could rule out radiative cooling as the only mechanism for causing the delay, since the rising part of the optical emission follows the rising part of the $\gamma$-rays. In addition, the decay time scale in the UVOT data seems comparable to that seen in the LAT, and not longer as required by energy-dependent dominated cooling. A reasonable mixture of SSC and an extra-contribution by Comptonization of the BLR photon field (external jet origin) by the same population of energetic electrons \citep[ERC, e.g.][]{sikora94,sikora01,dermer02,ghisellini09a} could be more adequate to explain also the SED, and such SSC+ERC scenario is still compatible with the X-ray to $\gamma$-ray correlation found in PKS 1502+106.
%
%
%
%
\par As seen in Section \ref{kanataresults}, the first of the fast, minor $\gamma$-ray flares occurring in Sept.~2008 (DoY 248-251) after the big outburst, was well observed by the Kanata-TRISPEC telescope in both $J$ and $V$ bands, with no significant time lag. The lower intensity and duration of the flare, the limited temporal resolution of the data, and a possible dominance of the SSC process during this episode can explain this difference with respect to August's big outburst. The match between the MeV-GeV and optical-near-IR flare, was also crucial for the firm identification with PKS 1502+106 (Sect. \ref{subsec:association}), and was a reasonable confirmation of the source-intrinsic nature of this variation seen in the LAT light curve (Fig.~\ref{fig:LATlightcurve} and Fig.~\ref{fig:OR103multifreq}). These $V$ and $J$ flux measurements, obtained after the conclusion of the outburst phase but on a period longer than the \textit{Swift} monitoring (Fig.~\ref{fig:OR103multifreq}), appear well correlated, though the $V-J$ color was less pronounced than the multi-band near-IR
colors reported in \citet{watanabe04} and Fig.~\ref{fig:spitzerIRS}, while the degree of optical polarization remained almost constant.
%
%
%
\subsection{Spectral energy distribution}\label{subsect:SED}

Variability is a powerful diagnostic to investigate blazar physics, but represents also a supplementary problem for the analysis of broad band spectral energy distributions (SED), where model
constraints are provided by simultaneous and well time-resolved multifrequency data. In the previous section we indicated that dominant synchrotron and SSC (in-jet) mechanisms can explain the radio-to-X-ray emission and correlations, whereas the origin of the high-power MeV-GeV bolometric emission could be better constrained by external Comptonization of the radiation from the broad line region (BLR),
as invoked by \citep{ghisellini09b}, and found already in similar FSRQs of the EGRET era \citep[e.g.,][]{sokolov05,sikora08,sikora09}. The featureless mid-IR continuum observed by Spitzer-IRS in 2005 (Fig.~\ref{fig:spitzerIRS}), supports the hypothesis of a prominent synchrotron emission by the jet, controlling the lower energy component. It could also reflect the lack of detectable ambient dust radiation, thereby supporting the idea that the ERC dissipation occurred within the BLR for this LAT outburst, in agreement with prescriptions of \citep{sikora02}.
\par PKS 1502+106 might be considered a blazar peaked at the border of the MeV and GeV bands (a peak around 0.4-0.5 GeV is suggested by the curved model fit of the 4-month spectrum, Section \ref{subsect:gamma-spectra}). In other words, PKS 1502+106 is likely at the border of the family of BLR-dissipated FSRQs and circum-nuclear ambient/torus dust-dissipated FSRQs, with an important SSC power output from radio-to-X-ray bands (Section \ref{subsect:gamma-x-uv-correlation}), as depicted by the multiband correlations, the absence of hints for a bulk Compton feature produced by cold, adiabatically cooling electrons \citep[as observed in PKS 1510-089, ][]{kataoka08,pks1510LAT}, the lack of evidence for a blue bump.

\par The medium or high black hole mass of this blazar \citep[likely in the range $ 0.5-1 \times 10^{9}$ M$_\sun$, as calculated by ][ respectively]{delia03,liu06}
could connote a BLR radiation field and inverse Compton dissipation moderately stronger than the magnetic energy density and the SSC luminosity in the gamma-rays, as reported
in Fig.~\ref{fig:SEDLAT2008}. The estimated accretion rate is 2 M$_{\sun}$ yr$^{-1}$ \citep{delia03}. In this case we have $L_{ERC}/L_{syn}=U'_{BLR}/U_{B}\simeq L_{BLR} \mathcal{D}^2 / (4\pi c R_{BLR}^{2} U_{B})$ where $U'_{BLR}$ and $U_{B}$ are the electromagnetic energy density in the BLR and the magnetic field energy density in the jet blob, respectively. With a $10^{9}$ M$_\sun$ mass the Schwarzschild radius of the supermassive black hole is $R_{S} = 2GM/c^{2} \simeq 3 \times 10^{14}$ cm. The bolometric luminosity of the BLR in PKS 1502+106, evinced by the Mg\texttt{II} emission line profile, is $L_{BLR} = 3.7 \times 10^{45}$ erg s$^{-1}$ \citep{liu06}, implying that the BLR is located at a radius $R_{BLR}= 10^{17} \sqrt{L/10^{45}} \simeq 2 \times 10^{17}$ cm. We can expect relatively smaller magnetic fields in the dissipation region and an ERC, external-jet process that is the dominant cooling mechanism when the source is in a high $\gamma$-ray state, as suggested in Section \ref{subsect:gamma-variability}.
In fact a SSC-only model would require a magnetic field in very sub-equipartition conditions, even if only the BLR fraction lying within
the varying region's beaming cone could contribute to gamma-rays. The simultaneous SED corresponding to the outburst state reported in Fig.~\ref{fig:SEDLAT2008} corroborates the possible needs for an ERC contribution, showing a large ($L_{\gamma}/L_{opt} \sim 100  $) gamma-ray dominance over the synchrotron component. The SSC+ERC model can be considered in some way in agreement with the redshift value of this blazar, if we assume the FSRQs~$\to$~BL~Lacs cosmological evolutionary scenario with dimming jet power, possibly related to star formation rate or the the far-IR/submillimeter luminosity density \citep{dermer07}.
\par In Fig.~\ref{fig:SEDLAT2008} we report the data of the two SED states (the ``outburst state'' of Aug.~05-10, 2008 (blue points) and the ``post-flare'' state (red points), characterized by a intermediate luminosity, of Aug.~11-22, 2008 (both intervals are outlined in Fig.~\ref{fig:LATlightcurve}), in conjunction with the unpublished archival and literature data collected for comparison. Above the two simultaneous SEDs of Aug.~2008, SSC and ERC ``strawman'' models are reported for each state (SSC+ERC modeling, blue lines for the outburst period, and SSC+ERC plus a SSC-stand-alone modeling, red lines, both for the post-flare period).
\par The pure SSC stand-alone model for the post-flare state (indicated in Fig.~\ref{fig:SEDLAT2008} as a version 2 of the first order SSC modeling with the label ``SSC2'', continuous red line, spanning from radio to gamma-ray bands) implements the temporal evolution of the synchrotron and SSC spectral components in a single flaring blob within the jet, where a population of accelerated electrons having a power-law with exponential cutoff distribution ($dN/dE \propto E^{-p}e^{-(E/E_{max})} $), is instantly injected \citep[for more details on the analytical and numerical model see, for example, ][]{ciprini08}. This form for $dN/dE$ is plausible in presence of time-dependent acceleration or radiative-loss limits \citep{webb84,drury91}. The post-flare (lower) state is reproduced by setting the electron energy index $p=1.77$, minimal and maximal electron Lorentz factors $\gamma_{min}=200$, $\gamma_{max}=8 \times 10^{4}$, compactness injected $\ell_{inj} = L_{inj} \sigma_{T} / (R m_{e} c^{3} ) = 10^{-3}$, radius of the emitting region in the comoving frame $R=6.5 \times 10^{17}$ cm, a bulk Doppler factor $\mathcal{D}=8$ and magnetic field intensity $B=0.024$ G. If the X-ray spectrum is produced by electrons that cool on timescales longer than the light crossing time, $R/c$, the X-ray spectral index would be $\alpha_{X}=(p-1)/2\simeq 0.4$, a value quite similar to the averaged value ($\alpha_{X}=0.45\pm0.03$) measured by \textit{Swift} XRT for the post-flare interval (bottom segment of Table \ref{tab:swiftXRT2008}). The rather large size of the emitting region ($R\simeq 0.2$pc) and the very sub-equipartition magnetic field and reduced $\mathcal{D}$ suggested by this ``SSC2'' (stand-alone and first order radio-to-gamma-ray SSC) model attempt indicate that even for this lower brightness state, an ERC contribution can be reasonable. Furthermore, the value of the optical-UV spectral index for this post-flare state ($\alpha_{UVOT} =1.9\pm 0.3$) is softer than the averaged index (3/2) expected \citep{chiang02}, when a totally SSC-dominated loss is considered. The huge Compton dominance of the outburst state might preclude a first order SSC-stand-alone model attempt for such SED state. On the other hand for high-energy $\gamma$-ray loud blazars like PKS 1502+106 superquadratic variations produced by higher order scatters in SSC are predicted by other models \citep{georganopoulos06,perlman08} and Section \ref{subsect:gamma-variability}, in fact in luminous blazars, the optical depth to Compton scattering (related to particle density) increases and second and higher-order scatters become more important. In more Compton-dominated objects these scattering reactions can dominate the energy output from the SSC process, and can produce superquadratic behaviors during big flares.
\par As mentioned above a ``SSC plus ERC'' model description, joining a SSC radio-to-X-ray component and a ERC gamma-ray component, to both the post-flare and outburst states (Fig.\ref{fig:SEDLAT2008}, dashed blue and red dotted lines) is plausible as well for these SEDs. The description of an initial version of this composite modeling, implementing the prescriptions of \citet{sikora94,dermer02}, can be found, for example, in \citet{tramacere03}. The low state is described with the following ``SSC+ERC'' parameters $B = 0.5$ G, blob size $R = 7.9 \times 10^{16}$ cm, $\mathcal{D}=20$, a log-parabola electron injection function between $\gamma_{min} = 100 $ and $\gamma_{max} = 3 \times 10^{4}$, $L_{disk}= 1.1 \times 10^{46}$, $\tau_{BLR}=0.1$, $T_{disk}=1.5 \times 10^{5}$ $^{\circ}$K and distance from the disk of $10^{18}$ cm. The outburst state is described with the following ``SSC+ERC'' parameters $B = 0.5$ G, blob size $R = 6.3 \times 10^{16}$ cm, $\mathcal{D}=24$, a log-parabola electron injection function between $\gamma_{min} = 100 $ and $\gamma_{max} = 3 \times 10^{4}$, $L_{disk}= 1.1 \times 10^{46}$, $\tau_{BLR}=0.1$, $T_{disk}=1.5 \times 10^{5}$ $^{\circ}$K and distance from the disk of $7 \times 10^{17}$ cm.
\par In Fig.~\ref{fig:SEDLAT2008}, the LAT averaged spectrum (E$>$200 MeV) of the entire Aug.-Dec.~2008 period considered (characterized by high statistics and consistent with an intrinsically curved shape described by a log-parabola model) is reported as well (the strip in the same orange color used for non-simultaneous data). Log-parabola curvature (Section \ref{subsect:gamma-spectra}) at high energy can be produced by several plausible models. For example by radiative particle cooling and stochastic acceleration processes driven by magnetic turbulence (rather than systematic particle acceleration) acting near the shock front \citep{stawarz08}. If the energy where losses balance the acceleration rate if the acceleration
time decreases more slowly than the loss time. The probability of energetic gain is lower when particle energy increases, because particles are confined by a magnetic field with a confinement efficiency decreasing for an increasing gyration radius. The integral energy distribution of the accelerated particles results in a log-parabolic law:  $N(\gamma)=N_{0}(\gamma/\gamma_{0})^{-p-1+r \log (\gamma/\gamma_{0}) }$, where $r$ is a curvature term \citep{landau86,fossati00,massaro04,perlman05,tramacere07}.
Non-power law radiation spectra could also originate in a non-linear regime, such as in the presence of shock modifications, precursors, and when diffusion coefficients vary with particle momentum \citep{amato08}. Another explanation takes into account episodic particle acceleration, that is applicable to high-energy flares with intrinsic spectral curvature \citep{perlman05,perlman05b}. The filling factor of the regions within which particles are accelerated is a function of both position and energy. If the light-crossing time of the emission region or the integration time of observations is greater than the
characteristic particle acceleration time, we effectively observe an electron distribution which is the product of a power law multiplied
by a logarithmic term producing a spectral curvature.
\par Another contribution to such deviations might also be the averaging over long periods of time, hence combining variability effects, resulting in a cumulation of photons from different activity ``flavors'' (different source brightness and spectral hardness) being included in the evaluation of the spectrum.
In the case of PKS 1502+106, the same spectral curved shape was also preferred for the much shorter, post-flare, interval, while the whole period includes many photons and covers a wider energy range (photons observed from the source have energies up to 15.8 GeV). PKS 1502+106, and 3C 454.3 \citep{3c454lat08} for the first time show this
departure from a simple power law, while EGRET observed only simple power
laws, probably due to its lower high-energy observational limit. The maximum
peak observed in the LP models of PKS 1502+106 (in the $\nu F_{\nu}$
representation) are around energies of about 390 MeV for the whole period
and 480 MeV for the post-flare period. These peaks are consistent with a sub-GeV FSRQ blazar class.
\par From Sect. \ref{subsect:gamma-variability} we remember that the inferred, apparent and isotropic, monochromatic luminosity at $E_{0}=100$ MeV during the outburst phase (DoY 2008: 218.95 - 224.0) is $L_{E>100} \simeq 1.1 \times 10^{49} $ erg s$^{-1}$, and the bolometric luminosity is expected to be even higher than this value, making PKS 1502+106 one of the most powerful and luminous high-energy blazars observed during the first year the \textit{Fermi} LAT all-sky survey. On the other hand The highest energy of photons detected from the source during the first four months of survey (15.8 GeV, assuming a strict PSF size criterium), has marginal consequences for extragalactic background light (EBL) predictions. The optical depth for $\gamma\gamma \to e^{+}e^{-}$ pair production of 16 GeV photons propagating through the EBL from a redshift $z=1.839$ source to Earth approaches unity for rather high-density EBL models \citep[e.g, ][$\tau_{\gamma\gamma}(z=1.839, E=16~\mathrm{GeV}) \sim 1.0-1.3$]{stecker06}, while most low density EBL models predict rather small interaction probabilities at such energies. An in-depth exploration of this finding  will be presented elsewhere \citep{ebl_lat}.
%
%
%
%
\section{Final remarks}\label{sect:conclusion}
%
%
This was the first time that PKS 1502+106, a distant radio- and X-ray-selected FSRQ, has been announced to have observable high-energy gamma-ray emission above 100 MeV. \textit{Fermi} LAT as basically a sort of all-space, -time, and -energy monitor allowed an excellent spatial localization of this new $\gamma$-ray source (Section \ref{subsec:association} and Fig.\ref{fig:countmap}, with a firm identification thanks to the optical-$\gamma$-ray match of the outburst and, remarkably, also of a second flare). It allowed detailed analysis of the energy spectrum (Section \ref{subsect:gamma-spectra} and Fig. \ref{fig:SEDLAT2008}, demonstrating the possibility for an intrinsic spectral curvature in $\gamma$-rays), and was able to provide regular, daily monitoring flux light curve (Section \ref{subsect:gamma-lightcurve} \ref{fig:LATlightcurve}). This has made possible the discovery of consistent $\gamma$-ray brightness and activity ignited by the big outburst over the following 4 months (pointing out a $1/f^{1.3}$ variability behavior), the discovery of further minor and rapid flares, and the disclosure of the outburst's temporal shape. This type of ``PSD-SED'' monitoring performed by the LAT yielded advantage in developing the first unplanned \textit{Fermi} multifrequency campaign, with a strategic 16 days of simultaneous \textit{Fermi-Swift} monitoring.
\par PKS 1502+106 is a powerful gamma-ray ($\sim 10^{49}$ erg s$^{-1}$ at $E>100$ MeV) FSRQ that showed, especially during the fast-rising outburst, a dominant MeV-GeV bolometric emission similar to other FSRQs of the EGRET era. Dissipation probably occurred within the BLR, and, assuming for a black hole mass of $\sim 10^{9}$ M$_\sun$, the $\gamma$-ray emission was likely dominated by the ERC process. The SSC, in-jet, emission appears to dominate the observed SED from radio to X-rays bands. PKS 1502+106 might be considered an example of a sub-GeV peaked blazar, placed at the border of the BLR dissipated and the dusty torus/ambient-radiation dissipated FSRQs classes. The level of correlation found among the $\gamma$-ray, X-ray and optical-UV outburst and post-flare relaxing phases, support this idea. Opacity effects at cm and longer radio wavelengths, possible links between field ordering, jet-axis alignment, superluminal radio knots, and MeV-GeV outburst, are also depicted by our results.
\par In conclusion, the \textit{Fermi} LAT performance in blazar science (as a stand alone observatory, or leading instrument for multifrequency campaigns), and the synergy between \textit{Fermi} and \textit{Swift} in particular, is evidenced by this work on PKS 1502+106. By itself, this blazar is emerging as a major, luminous, energetic and $\gamma$-ray variable source, with promising diagnostic and discovery potential in emission modeling, in spectral and temporal variability studies, and in understanding the radio-gamma-ray connection.
%
%
%

%
\section{Acknowledgments}
%
%
This research is based on observations obtained with the \textit{Fermi}
Gamma-ray Space Telescope. The \textit{Fermi} LAT Collaboration acknowledges generous ongoing support from a number of agencies and institutes that have supported both the
development and the operation of the LAT as well as scientific data analysis.
These include the National Aeronautics and Space Administration and the
Department of Energy in the United States, the Commissariat \`a l'Energie Atomique
and the Centre National de la Recherche Scientifique / Institut National de Physique
Nucl\'eaire et de Physique des Particules in France, the Agenzia Spaziale Italiana (ASI)
and the Istituto Nazionale di Fisica Nucleare (INFN) in Italy, the Ministry of Education,
Culture, Sports, Science and Technology (MEXT), High Energy Accelerator Research
Organization (KEK) and Japan Aerospace Exploration Agency (JAXA) in Japan, and
the K.~A.~Wallenberg Foundation, the Swedish Research Council and the
Swedish National Space Board in Sweden.
\par %
Additional support for science analysis during the operations phase is gratefully
acknowledged from the Istituto Nazionale di Astrofisica (INAF) in Italy and the Centre National d'\'Etudes Spatiales in France.
\par %
SC acknowledges funding by grant ASI-INAF n.I/047/8/0 related to
Fermi on-orbit activities.
\par %
This research has made use of the NASA/IPAC NED
database (JPL CalTech and NASA, USA), the HEASARC database (LHEA
NASA/GSFC and SAO, USA), the Smithsonian/NASA's ADS bibliographic
databases, and the SIMBAD database (CDS, Strasbourg, France). This
work includes observations obtained with the NASA \textit{Swift}
gamma-ray burst Explorer. This work includes observations
obtained with the Spitzer Space Telescope (operated by the Jet
Propulsion Laboratory, California Institute of Technology under a
contract with NASA). This work includes observations
obtained with XMM-Newton, an ESA science mission with instruments
and contributions directly funded by ESA Member States and NASA. This
work has made use of observations obtained with Owens Valley Radio Observatory. The monitoring program at the OVRO is supported by NASA award NNX08AW31G, and NSF award\# AST-0808050.   This research has made use of observations from the MOJAVE database that is maintained by the MOJAVE team. The MOJAVE project is supported under National Science Foundation
grant 0807860-AST and NASA-\textit{Fermi} grant NNX08AV67G. The National
Radio Astronomy Observatory (NRAO VLBA) is a facility of the
National Science Foundation operated under cooperative agreement
by Associated Universities, Inc.   This research has made use of observations obtained with the 100-m telescope of the MPIfR (Max-Planck-Institut f\"ur Radioastronomie)
at Effelsberg, Germany.    This research has made use of
observations from the RATAN--600 that is partly supported by the
Russian Foundation for Basic Research (projects 01-02-16812,
05-02-17377, 08-02-00545). This work has made use of observations obtained with the
14m Mets\"{a}hovi Radio Observatory, a separate research institute of the Helsinki University of Technology. The
Mets\"{a}hovi team acknowledges the support from the Academy of Finland.
This work has made use of observations obtained with the
TRISPEC instrument on the Kanata telescope that is operated by Hiroshima University,
Japan. YYK is a Research Fellow of the Alexander von Humboldt Foundation.
\par %
The LAT team and multifrequency collaboration extend thanks to the anonymous referee who
made very useful comments.\\

{\it Facilities:} \facility{{\it Fermi} LAT}.

\bibliographystyle{apj}

\begin{thebibliography}{}


\bibitem[Abdo et al.(2009a)]{3c454lat08} Abdo, A.~A., et al., 2009a, \apj, 699, 817

\bibitem[Abdo et al.(2009b)]{3monthsagn} Abdo, A.~A., et al., 2009b, \apj, 700, 597

\bibitem[Abdo et al.(2009c)]{3monthcatalog} Abdo, A.~A., et al., 2009c, \apjs, 183, 46

\bibitem[Abdo et al.(2009d)]{6monthvariability} Abdo, A.~A., et al., 2009d, \apj,
submitted, (AGN variability)

\bibitem[Abdo et al.(2009e)]{pks1454foschini} Abdo, A.~A., et al., 2009e, \apj, 697, 934

\bibitem[Abdo et al.(2009f)]{pks1510LAT} Abdo, A.~A., et al., 2009f, \apj, submitted, (PKS 1510-089)

\bibitem[Abdo et al.(2009g)]{ebl_lat} Abdo, A.~A., et al., 2009g, \apj, in prep. (EBL study)

\bibitem[Abdo et al.(2009h)]{lat-calibration} Abdo, A.~A., et al., 2009h, Atropart. Physics, 32, 193

\bibitem[Aharonian et al.(2007)]{aharonian07} Aharonian, F., et
al.\ 2007, \apjl, 664, L71

\bibitem[Akiyama et al.(2003)]{akiyama03} Akiyama, M., Ueda, Y.,
Ohta, K., Takahashi, T., \& Yamada, T.\ 2003, \apjs, 148, 275

\bibitem[Aller et al.(1999)]{aller99} Aller, M.~F., Aller,
H.~D., Hughes, P.~A., \& Latimer, G.~E.\ 1999, \apj, 512, 601

\bibitem[Amato et al.(2008)]{amato08} Amato, E., Blasi, P.,
\& Gabici, S.\ 2008, \mnras, 385, 1946

\bibitem[An et al.(2004)]{an04} An, T., Hong, X.~Y., Venturi, T., Jiang, D.~R., \& Wang, W.~H.\ 2004, \aap, 421, 839

\bibitem[Angelakis et al.(2008)]{angelakis08} Angelakis, E.,
Fuhrmann, L., Marchili, N., Krichbaum, T.~P., \& Zensus, J.~A.\
2008, Memorie SAIt, 79, 1042

\bibitem[Argue \& Sullivan(1980)]{argue80} Argue, A.~N., \& Sullivan,
C.\ 1980, The Observatory, 100, 152

\bibitem[Atwood et al.(2009)]{michelsonLAT08} Atwood, W.~B., et al., 2009, \apj, 697, 1071

\bibitem[Atwood et al.(2007)]{atwood07} Atwood, W.~B., Bagagli, R.,
Baldini, L., et al.\ 2007, Astropart. Phys., 28, 422

\bibitem[Baars al.(1977)]{baars77} Baars, J.~W.~M., Genzel, R. Pauliny-Toth, I.~I.~K.,
Witzel, A., 1977, A\&A, 61, 99

\bibitem[Begelman et al.(1984)]{begelman84} Begelman, M.~C.,
Blandford, R.~D., \& Rees, M.~J.\ 1984, Reviews of Modern Physics, 56, 255


\bibitem[Bellazzini et al. (2002)]{bellazzini02} Bellazzini, R., et al. 2002, Nucl. Phys. B Proc. Sup. 113, 303

\bibitem[Bertin \& Arnouts(1996)]{bertin96} Bertin, E., \& Arnouts, S.\ 1996, \aaps, 117, 393

\bibitem[Blake(1970)]{blake70} Blake, G.~M.\ 1970, \aplett, 6, 201

\bibitem[Burbidge \& Strittmatter(1972)]{burbidge72} Burbidge, E.~M., \& Strittmatter, P.~A.\ 1972, \apjl, 174, L57

\bibitem[Burnett(2007)]{burnett07} Burnett, T.~H.\ 2007, AIP Conf.
Proc., 921, 530

\bibitem[B{\"o}ttcher
\& Chiang(2002)]{boettcher02} B{\"o}ttcher, M., \& Chiang, J.\ 2002, \apj, 581, 127

\bibitem[Camilo et al. (2009)]{camilo09} Camilo, F., Ray, P.~S., Ransom, S. M. et al. \ 2009, \apj, 705, 1

\bibitem[Casandjian \& Grenier(2008)]{casandjian08} Casandjian, J.-M., \&
Grenier, I.~A.\ 2008, \aap, 489, 849

\bibitem[Cecchi et al.(2007)]{cecchi07} Cecchi, C., Germani, S.,
Pepe, M., et al. 2007, AIP Conf. Proc., 921, 540

\bibitem[Chiaberge \& Ghisellini(1999)]{chiaberge99} Chiaberge, M., \& Ghisellini, G.\ 1999, \mnras, 306, 551

\bibitem[Chiang et al.(2007)]{chiang07} Chiang, J., Carson, J.,
\& Focke, W.\ 2007, AIP Conf. Proc., 921, 544

\bibitem[Chiang et al.(2006)]{chiang06} Chiang, J., Digel, S.,
Silva, E.~D.~C.~E., \& Reimer, O.\ 2006, Bull. American Astron.
Soc., 38, 382

\bibitem[Chiang \& B\"{o}ttcher(2002)]{chiang02} Chiang, J.,
B\"{o}ttcher, M.\ 2002, \apj, 564, 92

\bibitem[Ciaramella et al.(2004)]{ciaramella04} Ciaramella, A., Bongardo, C., Aller, H.D. et al.\ 2004, \aap, 419, 485

\bibitem[Ciprini(2008)]{ciprini08} Ciprini, S.\ 2008, in proc. of Blazar Variability across the Electromagnetic Spectrum, PoS(BLAZARS2008), n.073

\bibitem[Ciprini et al.(2007a)]{ciprini07a} Ciprini, S., Tosti, G.,
Marcucci, F. et al. 2007a, AIP Conf. Proc., 921, 546

\bibitem[Ciprini et al.(2007b)]{ciprini07b} Ciprini, S., Takalo, L. O., Tosti, G.,  et al.\ 2007b, \aap, 467, 465

\bibitem[Cooper et al.(2007)]{cooper07} Cooper, N.~J., Lister, M.~L., \& Kochanczyk, M.~D.\ 2007, \apjs, 171, 376

\bibitem[Crowther \& Clarke(1966)]{crowter66} Crowther, J.~H., \& Clarke, R.~W.\ 1966, \mnras, 132, 405

\bibitem[Damiani et al.(1997)]{damiani97} Damiani, F., Maggio, A., Micela, G., \& Sciortino, S. 1997, \apj, 483, 350

\bibitem[d'Arcangelo et al.(2007)]{darcangelo07} d'Arcangelo, et al. 2007, \apj, 659, L107

\bibitem[Day et al.(1966)]{day66} Day, G.~A., Shimmins, A.~J., Ekers, R.~D., \& Cole, D.~J.\ 1966, Australian Journ.
Phys., 19, 35

\bibitem[D'Elia et al.(2003)]{delia03} D'Elia, V., Padovani,
P., \& Landt, H.\ 2003, \mnras, 339, 1081

\bibitem[Dermer \& Schlickeiser(2002)]{dermer02} Dermer, C.~D., \& Schlickeiser, R.\ 2002, \apj, 575, 667

\bibitem[Dermer(2007)]{dermer07} Dermer, C.~D.\ 2007, \apj, 659, 958

\bibitem[Drury(1991)]{drury91} Drury, L.~O.\ 1991, \mnras, 251,
340

\bibitem[Fichtel et al.(1994)]{fichtel94} Fichtel, C. E., Bertsch, D. L., Chiang, J., et al. 1994, \apjs, 94,
551

\bibitem[Fitch et al.(1969)]{fitch69} Fitch, L.~T., Dixon,
R.~S., \& Kraus, J.~D.\ 1969, \aj, 74, 612

\bibitem[Foschini et al.(2006)]{foschini06} Foschini L., Ghisellini G., Raiteri C.M., et al., 2006, \aap 453, 829

\bibitem[Fossati et al.(2000)]{fossati00} Fossati, G., Celotti, A., Chiaberge, M., et al.\ 2000, \apj, 541, 166

\bibitem[Franceschini et al.(2008)]{franceschini08} Franceschini, A., Rodighiero, G., \& Vaccari, M.\ 2008, \aap, 487, 837

\bibitem[Fuhrmann et al.(2007)]{fuhrmann07} Fuhrmann, L., Zensus,
J.~A., Krichbaum, T.~P., Angelakis, E., \& Readhead, A.~C.~S.\
2007, AIP Conf. Proc., 921, 249

\bibitem[Fuhrmann et al.(2008)]{fuhrmann08} Fuhrmann, L., Krichbaum, T. P., Witzel, A. et al.\ 2008, \aap, 490, 1019

\bibitem[Gehrels et al.(2004)]{gehrels04}Gehrels, N., et al. 2004, \apj, 611, 1005

\bibitem[George et al.(1994)]{george94} George, I.~M., Nandra, K., Turner, T.~J., \& Celotti, A.\ 1994, \apjl, 436, L59

\bibitem[Georganopoulos et al.(2006)]{georganopoulos06} Georganopoulos,
M., Perlman, E.~S., Kazanas, D., \& Wingert, B.\ 2006,  AIP Conf. Proc., 350, 178

\bibitem[Georganopoulos \& Marscher(1998)]{georganopoulos98} Georganopoulos, M., \& Marscher, A.~P.\ 1998, \apjl, 506, L11

\bibitem[Ghisellini \& Tavecchio(2009a)]{ghisellini09a} Ghisellini, G., \& Tavecchio, F.\ 2009a, \mnras, 397, 985

\bibitem[Ghisellini et al.(2009b)]{ghisellini09b} Ghisellini, G.,
Tavecchio, F., \& Ghirlanda, G.\ 2009b, accepted, \texttt{arXiv:0906.2195}

\bibitem[Graff et al.(2008)]{graff08} Graff, P.~B.,
Georganopoulos, M., Perlman, E.~S., \& Kazanas, D.\ 2008, \apj, 689, 68

\bibitem[Hartman et al.(2001)]{hartman01} Hartman, R.~C., B\"{o}ttcher, M.,
Aldering, G.,  et al.\ 2001, \apj, 553, 683


\bibitem[Hartman et al.(1999)]{hartman99} Hartman, R.~C., Bertsch, D. L., Bloom, S. D.,
et al.\ 1999, \apjs, 123, 79

\bibitem[Healey et al.(2008)]{healey08} Healey, S. E., Romani, R. W.,
Cotter, G., Michelson, P. F., Schlafly, E. F., Readhead, A. C. S.,
Giommi, P., Chaty, S., Grenier, I. A.; \& Weintraub, L. C. 2008,
\apjs, 175, 97

\bibitem[Houck et al.(2004)]{houck04} Houck, J. R., et al. 2004, \apjs, 154, 18

\bibitem[Hovatta et al.(2009)]{hovatta09} Hovatta, T., Valtaoja, E., Tornikoski, M., L{\"a}hteenm{\"a}ki, A.\ 2009, \aap, 494, 527

\bibitem[Hovatta et al.(2007)]{hovatta07} Hovatta, T., Tornikoski, M., Lainela, M., Lehto, H.~J., Valtaoja, E., Torniainen, I., Aller, M.~F., \& Aller, H.~D.\ 2007, \aap, 469, 899


\bibitem[Hughes et al.(1992)]{hughes92} Hughes, P.~A., Aller,
H.~D., \& Aller, M.~F.\ 1992, \apj, 396, 469

\bibitem[Inoue \& Takahara(1996)]{inoue96} Inoue, S., \& Takahara, F.\ 1996, \apj, 463, 555

\bibitem[Jorstad et al.(2007)]{jorstad07} Jorstad, S. et al. 2007, \aj, 134,

\bibitem[Kataoka et al.(2008)]{kataoka08} Kataoka, J., Madejski, G.,
Sikora, M.,  et al.\ 2008, \apj, 672, 787

\bibitem[Kellermann et al.(2004)]{kellermann04} Kellermann K.~I., Lister M.~L., Homan
  D.~C., et al., 2004, \apj, 609, 539

\bibitem[Kirk et al.(1998)]{kirk98} Kirk, J. G., Rieger, F. M., Mastichiadis, A. 1998, A\&A, 333, 452

\bibitem[Komatsu et al.(2009)]{komatsu08} Komatsu, E., Dunkley, J., Nolta, M. R., et al. 2009, \apjs, 180, 330

\bibitem[Kovalev et al.(2009)]{kovalev09} Kovalev, Y.~Y., Aller, H. D., Aller, M. F.,  et al.\
2009, \apjl, 696, L17


\bibitem[Kovalev et al.(2005)]{kovalev05} Kovalev, Y.~Y., Kellermann, K. I., Lister, M. L., et al.\
2005, \aj, 130, 2473

\bibitem[Kovalev et al.(1999)]{kovalev99} Kovalev, Y.~Y., Nizhelsky, N.~A.,
Kovalev, Yu.~A., Berlin, A.~B., Zhekanis, G.~V., Mingaliev, M.~G.,
\& Bogdantsov, A.~V.\ 1999, \aaps, 139, 545

\bibitem[Korolkov \& Parijskij(1979)]{korolkov79} Korolkov, D.~V.,
\& Parijskij, Yu.~N.\ 1979, Sky~Telesc., 57, 324

\bibitem[Landau et al.(1986)]{landau86} Landau, R., et al.\
1986, \apj, 308, 78

\bibitem[Leipski et al.(2008)]{leipski08} Leipski et al.\ 2009, \apj, 701, 891


\bibitem[Lister et al.(2009a)]{lister09a} Lister M. L., Aller, H. D.,
Aller, M. F., 2009, \aj, 137, 3718

\bibitem[Lister et al.(2009b)]{lister09b} Lister, M.~L., Homan,
D.~C., Kadler, M., Kellermann, K.~I., Kovalev, Y.~Y., Ros, E., Savolainen, T., \& Zensus, J.~A.\ 2009, \apjl, 696, L22


\bibitem[Lister \& Homan(2005)]{lister05} Lister M.~L., \& Homan D.~C., 2005, \aj,
  130, 1389

\bibitem[Liu et al.(2006)]{liu06} Liu, Y., Jiang, D.~R.,
\& Gu, M.~F.\ 2006, \apj, 637, 669

\bibitem[L{\'o}pez-Caniego et al.(2007)]{lopez07} L{\'o}pez-Caniego, M.,
Gonz{\'a}lez-Nuevo, J., Herranz, D., et al.\ 2007, \apjs, 170, 108

\bibitem[Lott et al.(2007)]{lott07} Lott, B., Carson, J.,
Ciprini, S., et al.\ 2007, AIP Conf. Proc., 921, 347


\bibitem[Madejski et al.(1996)]{madejki96} Madejski, G.,
Takahashi, T., Tashiro, M., et al. 1996, \apj, 459, 156

\bibitem[Marcucci et al.(2004)]{marcucci04} Marcucci, F., Cecchi, C., \& Tosti, G.,\ 2004,
Frascati Physics Ser. XXXVII, 285

\bibitem[Marscher et al.(2008)]{marscher08} Marscher A. P.; Jorstad, S. G.; d'Arcangelo, F. D.,  et al. 2008, Nature, 452, 966

\bibitem[Massaro et al.(2009)]{massaro08} Massaro, E., Giommi, P., Leto, C., et al. 2009, \aap, 495, 691

\bibitem[Massaro et al.(2004)]{massaro04} Massaro, E., Perri, M., Giommi, P., \& Nesci, R.\ 2004, \aap, 413, 489

\bibitem[Mattox et al.(1996)]{mattox96} Mattox, J.~R., Bertsch, D. L., Chiang, J., et al.\
1996, \apj, 461, 396


\bibitem[Mattox et al.(1993)]{mattox93} Mattox, J.~R., et al.\
1993, \apj, 410, 609

\bibitem[McEnery(2006)]{mcenery06} McEnery, J.\ 2006, ASP Conf. Proc. 350, 229

\bibitem[Michelson(2007)]{michelson07} Michelson, P.~F.\ 2007, AIP Conf. Proc., 921, 8

\bibitem[Moskalenko et al.(2003)]{moskalenko03} Moskalenko, I.~V.,
Jones, F.~C., Mashnik, S.~G., Ptuskin, V.~S., \& Strong, A.~W.\
2003, Int. Cosmic Ray Conf., 4, 1925

\bibitem[Neugebauer et al.(1986)]{neugebauer86} Neugebauer, G.,
Miley, G.~K., Soifer, B.~T., \& Clegg, P.~E.\ 1986, \apj, 308, 815

\bibitem[Nolan et al.(2003)]{nolan03} Nolan, P.~L., Tompkins,
W.~F., Grenier, I.~A., \& Michelson, P.~F.\ 2003, \apj, 597, 615

\bibitem[Ogle et al. et al.(2009)]{ogle08} Ogle, P., et al.\ 2009, in prep.

\bibitem[Pei(1992)]{pei92}Pei, Y. C.  1992, \apj, 395, 130

\bibitem[Perlman et al.(2008)]{perlman08} Perlman, E., Addison,
B., Georganopoulos, M., Wingert, B., \& Graff, P.\ 2008, in proc. of Blazar Variability across the Electromagnetic Spectrum, PoS(BLAZARS2008), n.009

\bibitem[Perlman et al.(2005)]{perlman05} Perlman, E.~S., Madejski, G., Georganopoulos, M. et al.\ 2005, \apj, 625, 727

\bibitem[Perlman \& Wilson(2005)]{perlman05b} Perlman, E.~S., \& Wilson, A.~S.\ 2005, \apj, 627, 140

\bibitem[Petrucci et al.(2002)]{petrucci02} Petrucci, P. O., Henri, G., Maraschi,
L., et al. 2002, \aap, 388, L5

\bibitem[Pian et al.(2009)]{pian08} Pian, E., et al., \ 2009, in
prep.

\bibitem[Poole et al.(2008)]{poole08}Poole, T. S., Breeveld, A. A., Page, M. J.,  et al.  2008, \mnras, 383, 627

\bibitem[Readhead et al.(1989)]{readhead89} Readhead, A.~C.~S.,
Lawrence, C.~R., Myers, S.~T., Sargent, W.~L.~W., Hardebeck,
H.~E., \& Moffet, A.~T.\ 1989, \apj, 346, 566

\bibitem[Ritz(2007)]{ritz07} Ritz, S.\ 2007, AIP Conf. Proc., 921, 3

\bibitem[Schlegel et al.(1998)]{schlegel98}Schlegel, D. J., Finkbeiner, D. P., \& Davis, M.  1998, \apjs,
500, 525


\bibitem[Sikora et al.(2009)]{sikora09} Sikora, M., Stawarz,
{\L}., Moderski, R., Nalewajko, K., \& Madejski, G.~M.\ 2009, \apj, 704, 38


\bibitem[Sikora et al.(2008)]{sikora08} Sikora, M., Moderski,
R., \& Madejski, G.~M.\ 2008, \apj, 675, 71

\bibitem[Sikora et al.(2002)]{sikora02} Sikora, M., B{\l}a{\.z}ejowski, M.,
Moderski, R., \& Madejski, G.~M.\ 2002, \apj, 577, 78

\bibitem[Sikora et al.(2001)]{sikora01} Sikora, M.,
B{\l}a{\.z}ejowski, M., Begelman, M.~C.,
\& Moderski, R.\ 2001, \apj, 554, 1

\bibitem[Sikora et al.(1994)]{sikora94} Sikora, M., Begelman,
M.~C., \& Rees, M.~J.\ 1994, \apj, 421, 153

\bibitem[Smith et al.(1977)]{smith77} Smith, H.~E., Burbidge,
E.~M., Baldwin, J.~A., Tohline, J.~E., Wampler, E.~J., Hazard, C.,
\& Murdoch, H.~S.\ 1977, \apj, 215, 427

\bibitem[Sokolov \& Marscher(2005)]{sokolov05} Sokolov, A., \& Marscher, A.~P.\ 2005, \apj, 629, 52

\bibitem[Sokolov et al.(2004)]{sokolov04} Sokolov, A., Marscher,
A.~P., \& McHardy, I.~M.\ 2004, \apj, 613, 725

\bibitem[Sowards-Emmerd et al.(2005)]{sowards05} Sowards-Emmerd,
D., Romani, R.~W., Michelson, P.~F., et al.\ 2005, \apj, 626, 95

\bibitem[Sowards-Emmerd et al.(2003)]{sowards03} Sowards-Emmerd,
D., Romani, R.~W., \& Michelson, P.~F.\ 2003, \apj, 590, 109

\bibitem[Starck \& Pierre(1998)]{starck98} Starck, J.-L., \& Pierre, M., A\&AS, 128, 397

\bibitem[Stawarz \& Petrosian(2008)]{stawarz08} Stawarz, {\L}., \& Petrosian, V.\ 2008, \apj, 681, 1725

\bibitem[Stecker et al.(2006)]{stecker06} Stecker, F.~W., Malkan,
M.~A., \& Scully, S.~T.\ 2006, \apj, 648, 774


\bibitem[Str\"uder L. et al.(2001)]{struder01} Str\"uder L., Briel U., Dennerl K., et al., 2001, A\&A, 365, L18

\bibitem[Thompson(2007)]{thompson07} Thompson, D.~J.\ 2007, AIP Conf. Proc., 921, 86

\bibitem[Thompson (2006)]{thompson06} Thompson, D.J.,  2006, ASP Conf. Ser., 350, 113

\bibitem[Thompson et al.(1996)]{thompson96} Thompson, D.~J., Bertsch, D. L., Dingus, B. L. et al.\ 1996, \apjs, 107, 227

\bibitem[Ter{\"a}sranta et al.(2005)]{terasranta05} Ter{\"a}sranta, H., Wiren, S.,
Koivisto, P., Saarinen, V., \& Hovatta, T.\ 2005, \aap, 440, 409

\bibitem[Ter\"{a}sranta et al.(1998)]{terasranta98} Ter\"{a}sranta, H.,  Tornikoski, M., Mujunen, A., et al.\ 1998, \aaps, 132, 305

\bibitem[Tosti(2007)]{tosti07} Tosti, G.\ 2007, AIP Conf. Proc., 921, 255

\bibitem[Tramacere et al.(2007)]{tramacere07} Tramacere, A., Massaro, F., \& Cavaliere, A.\ 2007, \aap, 466, 521

\bibitem[Tramacere \& Tosti(2003)]{tramacere03} Tramacere, A., \& Tosti, G.\ 2003, New Astron. Rev., 47, 697

\bibitem[Uemura et al.(2008)]{uemura08} Uemura, M., Sasada, M., Arai, A.,  et al.\
2008, in proc. of Blazar Variability across the Electromagnetic Spectrum, PoS(BLAZARS2008), n.070

\bibitem[Watanabe et al.(2005)]{watanabe05} Watanabe, M., Nakaya, H.,
Yamamuro, T., et al.\ 2005, \pasp, 117, 870

\bibitem[Watanabe et al.(2004)]{watanabe04} Watanabe, C., Ohta. K., Akiyama, M., \& Ueda, Y., 2004, \apj, 610, 128

\bibitem[Webb et al.(1984)]{webb84} Webb, G.~M., Drury, L.~O., \& Biermann, P.\ 1984, \aap, 137, 185

\bibitem[Wehrle et al.(1998)]{wehrle98} Wehrle, A., et al. 1998, \apj, 497, 178

\bibitem[Wilkes et al.(1983)]{wilkes83} Wilkes, B.~J., Wright, A.~E., Jauncey, D.~L.,
\& Peterson, B.~A.\ 1983, Proc. of Astronomical Society of
Australia, 5, 2

\bibitem[Williams et al.(1967)]{williams67} Williams, P.~J.~S.,
Kenderdine, S., \& Baldwin, J.~E.\ 1967, \memras, 70, 53

\bibitem[Wright et al.(1979)]{wright79} Wright, A.~E., Peterson,
B.~A., Jauncey, D.~L., \& Condon, J.~J.\ 1979, \apj, 229, 73



%
\end{thebibliography}

\end{document}